\documentclass[12pt,reqno]{amsart}
\usepackage{amsmath}%
\usepackage{amsfonts}%
\usepackage{amssymb}%
\usepackage{graphicx}
\usepackage{wrapfig, adjustbox}
\usepackage{caption}
\usepackage[foot]{amsaddr}
\usepackage{natbib}
\usepackage{apalike}
\usepackage{bbm}
\usepackage{enumerate}
\usepackage{xcolor}
\usepackage{subfig}
\usepackage{chngcntr}
\usepackage{url}
\usepackage{float}
\counterwithin{figure}{section}
\counterwithin{table}{section}
\usepackage{multirow}
\usepackage[margin=2.4cm]{geometry} 
\usepackage[misc]{ifsym}
\usepackage{accents}
\newtheorem{theorem}{Theorem}
\theoremstyle{plain}

\newtheorem{definition}[theorem]{Definition}

\newtheorem{remark}[theorem]{Remark}

\numberwithin{equation}{section}
\numberwithin{theorem}{section}

\newcommand{\T}{\intercal}

\usepackage[outdir=./Bilder/]{epstopdf}
\begin{document}
\title{On The Investment Strategies in Occupational Pension Plans}
\author[]{Frank Bosserhoff*, An Chen*, Nils S{\o}rensen*, Mitja Stadje*}
\address[]
{*Institute of Insurance Science, Ulm University, Helmholtzstrasse 20, 89081 Ulm, Germany}%
\email[]{\{frank.bosserhoff,an.chen,nils.sorensen,mitja.stadje\}@uni-ulm.de}%

\date{\today}
\keywords{Optimal Asset Allocation, Defined Contribution Plans, Target Date Funds}%
\thanks{Frank Bosserhoff acknowledges support from \textit{Deutscher Verein f\"ur Versicherungswissenschaften e.V.}}

\begin{abstract} 
 Demographic changes increase the necessity to base the pension system more and more on the second and the third pillar, namely the occupational and private pension plans; this paper deals with Target Date Funds (TDFs), which are a typical investment opportunity for occupational pension planners. TDFs are usually identified with a decreasing fraction of wealth invested in equity (a so-called glide path) as retirement comes closer, i.e., wealth is invested more risky the younger the saver is. We investigate whether this is actually optimal in the presence of non-tradable income risk in a stochastic volatility environment. The retirement planning procedure is formulated as a  stochastic optimization problem. We find it is the (random) contributions that induce the optimal path exhibiting a glide path structure, both in the constant and stochastic volatility environment. Moreover, the initial wealth and the initial contribution made to a retirement account strongly influence the fractional amount of wealth to be invested in risky assets. The risk aversion of an individual mainly determines the steepness of the glide path.
\end{abstract}
\maketitle

\section{Introduction}

One of the major societal challenges in various countries is the changing demographics in the sense of an aging society.
While birth rates remain low, life expectancy has been increasing continuously for several decades, leading to high costs for social security systems. Consequently, a shrinking working population has to support pensions of a growing retiring population, destroying an effective functioning of the statutory pay-as-you-go system. Hence, the role of the second and the third pillar, namely the occupational and private pension plans, is expected to gain more and more importance in the future. This paper is concerned with mechanisms ensuring a functioning second pillar.\par
Traditionally, Defined Benefit (DB) plans were the leading form of occupational retirement plans in developed countries. In a DB plan, the employee's pension benefit is determined by a formula taking into account years of service for the employer and wages or salary.
In the last decades, most industrial countries have experienced the conversion of DB plans to so-called Defined Contribution (DC) plans (see e.g. \cite{broeders2010pension}).
In a DC plan, sponsoring companies (and often also their employees) pay a promised contribution to an external pension fund which invests the contributions in financial assets. The pension payment is then simply determined as the market value of the backing assets. The fundamental advantages of DC plans are twofold. First of all, a DC plan allows pension beneficiaries to invest more freely and participate in the higher stock market returns which in particular in times of low interest rates seems necessary to obtain sufficient overall returns in the long run. Second, again in environments of falling interest rates, it is increasingly burdensome for DB pension funds or insurance companies to provide guarantees which require building up high and costly reserves in order to maintain them. Thus, in all OECD countries DC plans have played an increasingly important role, see also the discussion in \cite{aaronson2005firms} for further reasons for the transition from DB towards DC plans. Needless to say, in a DC plan the employees bear the entire investment risk. For instance, DC beneficiaries in the U.S. manage the investment risk by so called ``Individual Retirement Accounts'', or more frequently by making contributions to so-called 401(k) plans, see \cite{copeland2011target} for details on these plans. For a 401(k) DC beneficiary, one of the default investment strategies is a Target Date Fund (TDF). TDFs are investment funds with a pre-specified maturity (target date). This target date is usually the time point of retirement of some individual or a group of individuals. Because of their structure, these funds place themselves in the category of ``life-cycle'' funds, rather than in the category of ``life-style'' funds where the risk profile of the investor is the investment paradigm.
 These funds are the investment side of a DC plan and directly coupled with the planned retirement year of the DC plan beneficiaries. They have the advantage that the pension beneficiaries do not have to choose a number of investments but only a single fund. The main mechanism behind these TDFs is that those who retire later shall invest more in equity, while those who retire earlier shall invest less in equity. In other words, equity holdings in TDFs shall decrease in age. Therefore, TDFs are usually identified by practitioners with ``glide paths'', i.e., a {\sl decreasing} curve of the equity holding (as a fraction of wealth) over time. Assets in TDFs have grown from 71 billion U.S. dollars in 2005 to 2.2 trillion U.S. dollars at the beginning of 2020 (\cite{morningstar}). \par
In this paper we investigate whether the TDF is a reasonable choice for the individual in a realistic financial market with a stochastic volatility and random endowment environment. We consider assets more complex than the usual log-normal case by modeling the instantaneous variance of the risky asset through a stochastic process. This way, the empirically verified fact that asset prices exhibit stochastic volatility is accounted for (see \cite{tankov2003financial} and references therein). The resulting financial market is known as Heston Model (\cite{heston1993closed}). Moreover, we model the contribution the employee makes to the investment fund as stochastic process as well. Since we explicitly exclude perfect correlation between the latter and the financial market, the contribution risk is not tradeable. From a practical point of view, it is reasonable to consider random contributions due to unforeseen changes in wages or salary, unemployment or reduced working hours, which has for example been faced by many workers due to the covid pandemic; however, a perfect correlation to risky assets seems unrealistic. 
In such a stochastic environment, we approach then the question from a mathematical point of view in the sense that we formulate the retirement saving problem as a stochastic optimization problem; if TDFs are a reasonable choice, the solution to such a stochastic optimization problem will display the aforementioned glide path structure. We aim at finding the optimal investment strategy for each pension beneficiary with different risk aversion and income levels (leading to different contributions to the pension fund). \par
Undoubtedly, one of the core mathematical foundations of the current project is optimal dynamic asset allocation. The optimal consumption and asset allocation problem (utility maximization problem) in a continuous-time setting dates back to \cite{merton1969lifetime, merton1971optimum}, where analytic solutions for the case of utility functions exhibiting hyperbolic absolute risk aversion are obtained.
Merton's pioneering work has been extended in numerous directions.
A second core foundation to this project is Heston model which is an example of a so-called stochastic volatility model. Originally, this has been introduced in \cite{heston1993closed} for option pricing and as possible explanation of the volatility smile. An application of this stochastic volatility model to utility maximization is found e.g. in \cite{kallsen2010utility}, \cite{kraft2013dynamic}, and \cite{chen2018optimal}.

\par
There is some academic literature dealing with DC plans: \cite{cairns2006stochastic} consider the optimization problem under a power utility function which uses the plan member's salary as numeraire. A guarantee is incorporated by \cite{guan2014optimal}, who allow for stochastic interest rates and stochastic volatility while assuming that the salary risk is fully hedgeable by trading in the risky assets. The problem of maximizing the expected utility at the time of retirement using a constant elasticity of variance model for the stock price dynamics, which is essentially an extension of the classical geometric Brownian motion, is discussed in \cite{gao2009optimal}. The utility is measured in terms of mean-variance efficiency in \cite{guan2015mean} and a game-theoretic formulation is found in \cite{li2017equilibrium}. All these approaches to determining the optimal investment strategy are fundamentally different from ours as, from a mathematical point of view, we allow for additional randomness and incorporate non-tradable risk simultaneously. 
Working in a complex market environment comes at the expense of a three-dimensional value function, that is, the optimization problem induced is challenging from a technical point of view. We solve it by means of a Least-Squares Monte Carlo (LSMC) approach following \cite{denault2017dynamic}. We remark that the LSMC methods have not been widely applied in insurance, or specifically retirement planning problems. 
It allows us to explicitly investigate whether the glide path structure, that is characteristic for TDFs in practice, is still optimal when random volatility and/or not-tradeable salary risk are present. Moreover, the LSMC algorithm is very flexible in the sense that we can directly draw comparisons with the solution obtained in a corresponding constant volatility environment, i.e., the effect of the stochastic volatility on retirement planning is illustrated in detail.  \par
In our paper it is shown that in a stochastic volatility environment accounting for contribution risk the glide path structure is still optimal. More specifically,  we find it is the  (stochastic) contribution that causes the glide path structure. Particular attention is paid to comparisons of the constant and the stochastic volatility case; the respective strategies show a similar qualitative behaviour, but the variance of the optimal terminal wealth in the constant volatility model is higher than in the random volatility environment. The risk aversion of an individual mainly determines the steepness of the glide path, while the ratio of the initial value of the pension income account and the initial contribution level severely impact the fraction of total wealth invested in the risky asset. Moreover, we illustrate that the drift and volatility of the contribution process only have a minor impact on the optimal investment strategy; this implies that the same TDF can be used for individuals with different contribution parameters, which is an important fact from a practical point of view. \\
The structure of this paper is as follows: Section 2 introduces the stochastic volatility model for the financial market, the contribution process and the pension account. The latter is then used to formulate the stochastic optimization problem reflecting the retirement planning problem. Section 3 explains the numerical procedure applied in detail. Furthermore, a discussion of parameter values is provided. Section 4 contains our main results. It starts with a brief review of well-known stochastic optimization problems that are actually special cases of the problems ultimately considered in this paper. 
Section 5 succinctly summarizes our main findings. Table \ref{tbl:abbrv} contains abbreviations frequently used in the remainder of this paper.

\begin{table}[H]
\centering
\begin{tabular}{c| c}
	Abbreviation & Term \\ \hline \hline
	CE & Certainty Equivalent \\ \hline
	CRRA & Constant Relative Risk Aversion \\ \hline
	CVM & Constant Volatility Model \\ \hline
	CVMRCP & Constant Volatility Model Random Contribution Problem \\ \hline
	DC & Defined Contribution \\ \hline
	LSMC & Least-Squares Monte Carlo \\ \hline
	RRA & Relative Risk Aversion \\ \hline
	SVM & Stochastic Volatility Model \\ \hline
	SVMRCP & Stochastic Volatility Model Random Contribution Problem \\ \hline
	TDF & Target Date Fund \\ \hline
\end{tabular}
\caption{Table of Frequently Used Abbreviations}
\label{tbl:abbrv}
\end{table}

\section{Model Setup} 
\subsection{Financial Market}
Let $T>0$ be the deterministic retirement time point of some individual. Clearly, the optimal trading strategy of a retirement planner might vary considerably depending on the market model presumed. In the sequel, we consider a financial market modeling the volatility of the risky asset as a stochastic process as well as a model assuming the volatility parameter to be constant. To this end, we work on a fixed filtered probability space $(\Omega, \mathcal{F}, \mathbb{P}).$ This probability space is equipped with a three-dimensional standard Brownian motion $W = \left(W^{(1)},W^{(2)},W^{(3)}\right)^{\T}$. Let $(\mathcal{F}_{t})_{t \in [0,T]}$ be the right-continuous completion of the filtration generated by $W.$ On this space, both financial markets are defined. It is supposed that they are frictionless and two financial assets can be traded continuously at arbitrary quantities. As usual, the pension beneficiary aims at gainfully investing in these assets to maximize her wealth at the time point of retirement. In addition, it is assumed that a constant share of the pension beneficiary's labor income is continuously invested into the pension income account. As the contribution made to the investment fund is the quantity relevant for retirement planning, we directly consider the contribution as a stochastic process. We next introduce the two optimization problems considered in this paper. 

\subsection{The Stochastic Volatility Model}

For some deterministic interest rate $r \geq 0,$ denote the price process of the risk-free asset, for example a bank account, by $B = (B_{t})_{t \in [0,T]}$ such that $B$ solves
\begin{equation} \label{eq:riskless}
dB_{t} = r B_{t}\ dt, 
\end{equation}
with $B_{0} = 1$. Regarding the risky asset $S=(S_{t})_{t \in [0,T]},$ we assume an evolution according to the \emph{Heston Model} as originally suggested in \cite{heston1993closed}, i.e.
\begin{align}\label{eq:Stock_heston}
dS_{t} &= \mu S_{t} \ dt + \sqrt{\nu_{t}} S_{t}\ dW_{t}^{(1)},\\ 
d\nu_{t} &= \lambda(\theta - \nu_{t}) \ dt + \sigma_{\nu} \sqrt{\nu_{t}}\ \left(\rho_{S}\ dW_{t}^{(1)} + \sqrt{1-\rho_{S}^{2}}\ dW_{t}^{(2)} \label{eq:vol_heston} \right),
\end{align}
with $S_{0} > 0, \nu_{0} = x  > 0$. The process $\nu = (\nu_{t})_{t \in [0,T]}$ is the instantaneous variance of the stock and modeled as Cox-Ingersoll-Ross (CIR) process. The parameters $\mu, \lambda, \theta$ and $\sigma_{\nu}$ are positive constants; in this case, $\mu$ is the instantaneous drift of $S,$ the parameter $\lambda$ is the rate at which $\nu_{t}$ reverts to $\theta,$ which is the long-term variance of $S,$ and $\sigma_{\nu}$ is the volatility of the volatility (also known as 'vol of vol') and therefore determines the variance of $\nu_{t}$. Furthermore, we assume that the \textit{Feller condition} $2\lambda\theta > \sigma_v^2$ is satisfied ensuring $\nu_t>0$ for all $t.$ Since $W^{(1)}$ and  $W^{(2)}$ are by definition two independent standard Brownian motions, the instantaneous variance consists of two parts, one correlated and one independent to the risky asset. If $\rho_{S} = \pm 1, \nu$ is perfectly positively or negatively correlated to $S.$ In case $\rho_{S} = 0,$ $\nu$ is independent of $S.$ Note that the risk induced by the variance is fully hedgeable through trading in the stock only if $\rho_{S} \in \{-1,1\};$ since this is not the case, the market defined through \eqref{eq:riskless} - \eqref{eq:vol_heston} is \emph{incomplete}. We henceforth refer to this market as Stochastic Volatility Model (SVM). \par
Denote the aforementioned contribution process of some representative individual by $C=(C_{t})_{t \in [0,T]}$ and assume that its dynamics are given by

\begin{equation} \label{eq:contribution}
dC_{t} = \mu_{C}C_t \ dt + \sigma_{C}C_t \left(\rho_{C}\ dW_{t}^{(1)} + \sqrt{1-\rho_{C}^{2}} \ dW_{t}^{(3)} \right),
\end{equation}
with $C_{0} = c > 0,$ constants $\mu_{C} \in \mathbb{R}, \sigma_{C} > 0$ and a correlation coefficient $\rho_{C} \in (-1,1)$ allowing for the same interpretation as $\rho_{S}$ above. Hence, the contribution risk is correlated to the investment risk, however, the Brownian motion $W^{(3)}$ resembles shocks to the contributions being independent from any movements at the financial market; in particular, the contribution risk cannot be hedged through trading the risky asset. \par
In a DC pension plan, the changes in the individuals' retirement account stem from trading gains, changes in the underlying price processes as well as continuous contributions made over time. More precisely, denote by $\pi = (\pi_{t})_{t \in [0,T]}$ the \textit{proportion} of the total wealth invested by the fund manager in the stock $S$, accordingly $1-\pi$ displays the fraction of wealth put into the bank account. The allocation considered in this paper is given by $(1-\pi, \pi)$ and if not explicitly stated otherwise, we always refer to the fraction of total wealth invested whenever mentioning the trading strategy in the sequel. Denote the pension wealth process induced by $P^{\pi} = (P^{\pi}_{t})_{t \in [0,T]}.$ It is assumed to solve the SDE

\begin{align} \label{eq:wealth}
\begin{split}
dP^{\pi}_{t} &= P_{t}^{\pi} \pi_{t} \ \frac{dS_{t}}{S_{t}} + P_{t}^{\pi} (1-\pi_{t}) \ \frac{dB_{t}}{B_{t}} + C_{t} \ dt \\
&= \left(P_{t}^{\pi} r + P_{t}^{\pi} \pi_{t}\ (\mu - r) + C_{t}\right)\ dt + P_{t}^{\pi} \pi_{t} \sqrt{\nu_{t}}\ dW_{t}^{(1)}, 
\end{split}
\end{align}
$P^{\pi}_{0} = p > 0$. For $C_{t} \equiv 0$ for all $t,$ the wealth process would be self-financing; hence, the form of \eqref{eq:wealth} shows that the only cash injections stem from additional contribution income. Next, we define the set of admissible trading strategies; for notational convenience we write $\mathcal{R}_{+} := \mathbb{R}_{+} \times \mathbb{R}_{+} \times \mathbb{R}_{+}.$ Let $A\subseteq \mathbb{R}$  be a fixed given closed, convex set, which can express plausible trading restrictions. For instance $A=[0,1]$ corresponds to the prohibition of short selling and borrowing.
\begin{definition} \label{def_strat}
\begin{itemize}
	\item A progressively measurable trading strategy $\pi$ is called \emph{admissible} if it takes values in $A$, and if for any point $(t,p) \in [0,T) \times \mathbb{R}_{+},$ there exists a unique c\`adl\`ag adapted strong solution $P^{\pi}$ to \eqref{eq:wealth} starting from $p$ at $s=t$ fulfilling $\mathbb{E}[|P_{s}^{\pi}|^{2}]  < \infty$ as well as $P_{s}^{\pi} > 0$ for all $s > t$.  \\
	\item An admissible trading strategy $\pi$ in the form $\pi_{s} = \pi(s,P_{s}^{\pi},\nu_{s},C_{s})$ for some measurable function $\pi: [0,T] \times \mathcal{R}_{+} \to A$ is called \emph{Markovian feedback strategy}. \\
	\item We denote the set of admissible trading strategies of Markovian feedback type by $\Pi.$
\end{itemize}
\end{definition}

Unless stated otherwise, in the remainder of the paper we always implicitly refer to strategies of Markovian feedback type.\par Since DC pension plans typically pay out a lump-sum instead of annuities to the beneficiaries, we consider this case in the sequel. A natural target of the DC plan investor is then the maximization of the expected utility of the terminal wealth. Due to its analytical tractability we consider the \emph{power utility function}, that is
\begin{align} \label{eq:power_ut}
\begin{split}
U:\ &\mathbb{R}_{+} \to \mathbb{R},\\
& z \mapsto \frac{z^{1-\gamma}}{1-\gamma},\ \ \gamma \geq 0, \gamma \neq 1,
\end{split}
\end{align}
which assumes that the investor has a constant relative risk aversion (CRRA) given by $\gamma$. The long-term behavior of the economy suggests that the long-term relative risk aversion cannot strongly depend on wealth, see e.g. \cite{campbell2002strategic}; thus, as retirement planning is rather a long-term investment, using a utility function with CRRA is well-motivated economically.\par We write $\mathbb{E}_{t,p,x,c}[\cdot] = \mathbb{E}[\cdot | P_{t}^{\pi} = p, \nu_{t} = x, C_{t} = c]$ for the conditional expectation given the quadruple $(t,p,x,c) \in [0,T] \times \mathcal{R}_{+}$. The attractiveness of some trading strategy $\pi$ given the aforementioned quadruple is then evaluated using the following functional:

\begin{definition} \label{def_gain}
The \emph{gain function} $J: [0,T] \times \mathcal{R}_{+} \times \Pi \to \mathbb{R}_{+}$ is defined by
$$J(t,p,x,c,\pi) := \mathbb{E}_{t,p,x,c}[U(P_{T}^{\pi})].$$ \end{definition}

The natural goal of a retirement planner is the maximization of the gain function, i.e., the following stochastic optimization problem has to be solved:
\begin{equation*}
v(t,p,x,c) = \sup_{\pi \in \Pi} J(t,p,x,c,\pi). \tag{SVMRCP}
\end{equation*}
We call $v$ the associated value function and name the problem SVMRCP (Stochastic Volatility Model Random Contribution Problem). Consequently, a control $\pi^{\star}$ is called \emph{optimal} for SVMRCP if for the initial condition $(t,p,x,c) \in [0,T] \times \mathcal{R}_{+}$, it holds that $v(t,p,x,c) = J(t,p,x,c,\pi^{\star})$. 

\subsection{The Constant Volatility Model}
The focus of the paper is to analyze the suitability of TDFs for long-term retirement goals under realistic market conditions like stochastic volatility. To outline the impact of such a fluctuating volatility, a comparison with the constant volatility case will be drawn.
Replacing the random instantaneous volatility in \eqref{eq:Stock_heston} by the constant $\sigma_{S} > 0$ yields the following dynamics for the risky asset:
\begin{equation} \label{eq:S_cvm}
dS_{t} = \mu S_{t} \ dt + \sigma_{S}  S_{t}\ dW_{t}^{(1)},
\end{equation}
with $S_{0} > 0.$ Obviously, the solution to \eqref{eq:S_cvm} is the time-homogeneous geometric Brownian motion. The combination of the riskless asset $B$ described through \eqref{eq:riskless} and \eqref{eq:S_cvm} is henceforth called Constant Volatility Model (CVM).  \par
Assuming the random contribution is still described by \eqref{eq:contribution}, the portfolio process induced by the CVM is different from \eqref{eq:wealth} through the substitution of $\sqrt{\nu_t}$ by $\sigma_{S}$ only. Sticking to the notation $P^{\pi}$ for the portfolio process in the CVM, the definition of an admissible trading strategy in the CVM is readily deduced from Definition \ref{def_strat}. In particular, an admissible trading strategy of Markovian feedback type is given in the form $\pi_{s} = \pi(s,P_{s}^{\pi},C_{s})$ for some measurable function $\pi: [0,T] \times \mathbb{R}_{+} \times \mathbb{R}_{+} \to A.$ Note that $\pi_{s}$ does not depend on $\nu_{s}$ anymore. Using the power utility function \eqref{eq:power_ut} again, a suitably adapted form of the gain functional leads to the following stochastic optimization problem:
\begin{equation*}
v(t,p,c) = \sup_{\pi \in \Pi} J(t,p,c,\pi). \tag{CVMRCP}
\end{equation*}
By a slight abuse of notation, we again use $v$ as the associated value function and name the occurring problem CVMRCP (Constant Volatility Model Random Contribution Problem). Consequently, a control $\pi^{\star}$ is called \emph{optimal} for CVMRCP if, for the initial condition $(t,p,c) \in [0,T] \times \mathbb{R}_{+} \times \mathbb{R}_{+},$ it holds that $v(t,p,c) = J(t,p,c,\pi^{\star})$.

\begin{remark}
\begin{enumerate}[(i)]
	\item The CVM is obviously a special case of the SVM that is obtained by replacing the CIR process through a constant. The random contribution is assumed to be the same in both optimization problems. The fundamental difference in the optimization problems SVMRCP and CVMRCP is the omission of the random volatility in the latter. Consequently, the value function corresponding to the CVMRCP is reduced by one dimension. The inclusion of a third dimension in the value function can raise significant numerical problems that are discussed in the next section. 
	\item In practice, a target date fund is typically composed of a basket of risky assets, i.e., \eqref{eq:Stock_heston} or \eqref{eq:S_cvm} would be a multi-dimensional process and \eqref{eq:wealth} would change accordingly. However, as the focus of the paper is to study the impact of the realistic financial market setting like stochastic volatility and the untradable salary risk on the optimal pension asset allocation, we just consider one representative risky asset. In this sense, the focus of this paper is not the optimal allocation of money among a collection of risky securities as the point of retirement comes closer, but a change in the category of assets money is allocated to, namely from risky to  conservative assets. 
	\item In practice, the fraction of wealth invested in risky assets by a TDF amounts to 40-75 \% (\cite{morstar16}). In our analyses below we mostly obtain higher values. This stems from the omission of any shortselling and regulatory constraints which are typically present in practice. 
\end{enumerate}
\end{remark}

\section{Algorithm and Data}

\subsection{A Least-Squares Monte Carlo Approach}

It is a well known fact that (stochastic) optimization problems typically increase in terms of complexity, the more dimensions the corresponding value function has. As the CVMRCP is a particular case of the SVMRCP, if we can find a functioning numerical procedure which handles the SVMRCP properly, it can also be applied to the CVMRCP. It is not unusual in this context to resort to dynamic programming techniques that ultimately lead to solving a Hamilton-Jacobi-Bellmann (HJB) equation when approaching a stochastic optimization problem, see e.g. \cite{pham} for a comprehensive overview.  However, a major disadvantage of the HJB approach is the so-called \emph{curse of dimensionality}, i.e., the running time is increasing significantly when considering a higher-dimensional problem. According to \cite{broadie2004stochastic} and \cite{andreasson2019bias}, the HJB approach actually becomes impractical when two dimensions are exceeded. Noting that the SVMRCP has a three-dimensional value function gives rise to the consideration of alternatives such as Monte Carlo methods. The Least-Squares Monte Carlo (LSMC) approach to stochastic optimization problems is capable of handling multiple risk sources without suffering from the curse of dimensionality. In addition, this approach is useful when some flexibility regarding the dynamics of the underlying stochastic processes is required; this is particularly important for us as we consider for example constant and stochastic volatility. The ability of the LSMC technique to find the optimal control in an expected utility problem is shown in \cite{kharroubi2013numerical}, for background information on LSCM techniques see e.g. \cite{longstaff2001valuing} and \cite{clement2002analysis} and references therein. In order to solve the SVMRCP and the CVMRCP, we use an LSMC algorithm following \cite{denault2017dynamic} adapted to our setting. This algorithm is a simulation-and-regression procedure that regresses on decision variables (the admissible trading strategy) and the exogenous state variables (portfolio value, contribution and possibly the volatility). The algorithm is designed for discrete time models, so we decide on $N \in \mathbb{N}$ intermediate time steps per year from $t=0$ to $t=T$, i.e. $\Delta t = 1/N,$ and we observe the realizations of our stochastic processes at times $t\in \Xi:=\{0, \Delta t, 2 \Delta t, \dots,T\}$. The algorithm consists of three parts, namely the introduction of prerequisites, the backward step and the forward step. We write down the algorithm for the SVMRCP and remark that a reduction to the CVMRCP is immediate by dropping the stochastic volatility component. \\ 
\noindent \\
\vspace{3mm}
\underline{Prerequisites:}
\begin{enumerate}
	\item \textbf{Gridpoints for the share invested:} Let $n_{\pi} \in \mathbb{N}$ be the number of grid points and $\{\pi_{i}\}_{i=1}^{n_{\pi}}$ the corresponding grid of possible proportions of total wealth invested in the \textit{risky} asset. In our case we allocate the $n_{\pi}$ points uniformly on the set of admissible trading strategies $A = [\underaccent{\bar}{\pi}, \bar{\pi}], -\infty < \underaccent{\bar}{\pi} < \bar{\pi} < \infty.$ 
	\item \textbf{Generation of paths:} Let $n_{r} \in \mathbb{N}$ denote the number of realizations considered. Simulate and store $n_{r}$ possible paths of the discretized two-dimensional state process $\left\{\left(C^{(j)}_{t},\nu^{(j)}_{t}\right)_{t \in \Xi}\right\}_{j=1}^{n_r}$.  
	\item \textbf{Basis functions:} Define the vector of basis functions $B$, which in our case consists of monomials of the control $\pi$ and the state variables influencing the optimal strategy; mixed terms are included as well: 
	\begin{align*}
		&&B(\pi,c,\nu)&=[1 \quad  \pi \quad  \pi^{2} \quad  c \quad  c^{2}\quad \nu \quad  \nu^{2} \quad  \pi c \quad  \pi \nu \quad  c \nu]^{\intercal}.  
	\end{align*}
	\item \textbf{Grid of possible wealth levels:} A grid of possible wealth levels at each point in time has to be computed and stored. This grid should be wide enough to cover the range of realistic values of wealth, but not so large such that its simulation induces computational problems. 	To obtain in-between points and simultaneously account for the fact that bounds of the wealth levels become wider as time progresses due to the contribution and trading gains, a fixed amount of in-between points is impractical. Instead, we follow a suggestion in \cite{denault2017dynamic} for such kind of situations. Empirical evidence suggests that the following methodology leads to a realistic and efficient grid: simulate $n_r$ wealth paths using the values of the state process from Step 2 plugged in into \eqref{eq:wealth}, and invest the fixed share $\bar{\pi}$ in the risky asset. For each time point  $t \in \Xi \setminus \{0,T\},$  the $q_1$- and $1-q_2$-quantile are determined and chosen as the lower bound $P_{t,\min}$ and the upper bound $P_{t,\max}$, respectively. The quantiles $q_1,q_2$ can freely be chosen to fit the concrete problem.
    Further, define a \emph{fixed} step-size $\Delta P,$ which is calculated after the first iteration as follows: $$\Delta P:=\frac{P_{\Delta t,\max}-P_{\Delta t,\min}}{n_p},$$
	whereby $n_p$ is the number of in-between points after one time step; it is chosen by the user.	
	The number of grid points at a certain time point is given by $N_{p,t}=\lceil (P_{t,\max}-P_{t,\min})/\Delta P\rceil, t \in \Xi \setminus \{0,T\},$ where $\lceil \cdot \rceil$ denotes the ceil function. The grid points are then defined by $P_{t,k}=P_{t,\min}+k\Delta P$, $k \in \{0,1,\cdots, N_{p,t}\}$. Note that $P_{t,\max}$ is in general not a grid point. For time $t=0$ the only gridpoint is the initial wealth, i.e. $P_{0,0}=p$.

\end{enumerate}
\noindent \\
\vspace{3mm}
\underline{The Backward Step:}\\
For the ease of notation, the indices $k,i,j$ always refer to the corresponding point on the sets discussed above. Further, a subscript always denotes the dependence on a grid point while a superscript refers to a path/choice dependence.

For each $t = T-\Delta t$ to $0$ and for each $k=1$ to $N_{p,t}$ do:
\begin{enumerate}
	\item[STEP 1:] Given some portfolio value $P_{t,k}$ (this is the $k$-th possible value at time $t$), generate \textit{all} possible wealth levels at time $t+\Delta t$ by combining every possible path of the return with each allocation point (in total, there are $n_{\pi} \cdot n_{r}$ possibilities): \\
	\begin{equation}\label{eq:next_step}
	P_{t+\Delta t}^{(k,i,j)} := \mathcal{T} \left(P_{t,k},\pi_{i},c^{(j)}_t, \nu^{(j)}_t\right), \ \ i=1,\dots,n_{\pi},\ \text{and} \ j=1,\dots,n_{r}.
	\end{equation}
	The function $\mathcal{T}$ is the rebalancing function for the wealth process, i.e., in our case the corresponding Euler-Maruyama scheme applied to \eqref{eq:wealth}.
	\item[STEP 2:] For each $P_{t+\Delta t}^{(k,i,j)}$ generate a corresponding value of the value function $v_{t+\Delta t}^{(k,i,j)}$. Note that for each fixed $k\in \{1,\cdots, N_{p,t}\},$ there are exactly $n_{\pi} \cdot n_{r}$ combinations. 	
	\begin{enumerate}
		\item \emph{If $t = T-\Delta t$}, it is the final time that money is re-allocated. The value function $v_{T}$ simply coincides with the utility (of the terminal wealth): 
		$$v_{T}^{(k,i,j)} = u\left(P_{T}^{(k,i,j)}\right) = u\left(\mathcal{T} \left(P_{T-\Delta t,k},\pi_{i},c^{(j)}_{T-\Delta t}, \nu^{(j)}_{T-\Delta t}\right)\right).$$
		\item \emph{If $t < T-\Delta t$}, the value $v_{t+ \Delta t}^{(k,i,j)}$ is computed by interpolation. The rationale is as follows: Since the collection of pairs $\{\bar{v}_{t+\Delta t,k}^{(j)}, P_{t+\Delta t,k}\}_{k=1}^{N_{p,t+\Delta t}}$ is known from STEP 6 (see below) in the previous time step, interpolation is used as follows (for fixed $j$ and $i$): 
		For each $P_{t+\Delta t}^{(k,i,j)}$	estimate $v_{t+ \Delta t}^{(k,i,j)}$ by linear interpolation on the set $\left\{\bar{v}_{t+\Delta t,k}^{(j)}, P_{t+\Delta t,k}\right\}_{k=1}^{N_{p,t+\Delta t}}$. During the interpolation,  $\bar{v}_{t+\Delta t,k}^{(j)}$ is transformed by the inverse utility function $u^{-1}(x)=((1-\gamma) x)^{1/(1-\gamma)}$. It is common to use transformation functions before and after interpolation to reduce a potential interpolation bias, see e.g. \cite{andreasson2019bias}, \cite{carroll1988transformation} and in particular \cite{denault2017dynamic} for our case. 
	\end{enumerate}
	\item[STEP 3:] To each combination $\left(\pi_{i},c^{(j)}_t, \nu^{(j)}_t\right),$ associate the basis vector $B^{(i,j)}_t:=B\left(\pi_{i},c^{(j)}_t, \nu^{(j)}_t\right)$. Note that there are $n_{r} \cdot n_{\pi}$ such combinations.
	\item[STEP 4:] Regress the dependent values $v_{t+\Delta t}^{(k,i,j)}$ on the independent basis vectors $B^{(i,j)}_t$ and obtain the corresponding vector of regression coefficients by $\beta_{t,k}$:
	\begin{align}\label{eq:lsqmethod}
	\beta_{t,k}=\arg\min_{\beta \in \mathbb{R}^{10}} \sum_{j=1}^{n_\pi}\sum_{i=1}^{n_r}\left[\beta^{\intercal} B\left(\pi_{i},c^{(j)}_t, \nu^{(j)}_t\right)-v_{t+\Delta t}^{(k,i,j)}\right]^2.
	\end{align}
	\item[STEP 5:] Optimize the regression surface w.r.t. the position $\pi$: for each $j=1,\dots, n_{r}$, the value $(c^{(j)}_t, \nu^{(j)}_t)$ is known, so the following quantity gives the optimal portfolio weight:
	\begin{align}\label{eq:hatpi}\hat{\pi}_{t,k}^{(j)} = \arg\max_{\pi \in A}\beta_{t,k}^{\intercal}\ B\left(\pi, c^{(j)}_t, \nu^{(j)}_t\right), \ \ j=1,\dots, n_{r}. \end{align}
	This means that for every wealth level $P_{t,k},$ we have a collection $\left\{\hat{\pi}_{t,k}^{(j)}\right\}_{j=1}^{n_{r}}$ of optimal portfolio weights. 
	\item[STEP 6:] Compute the values $\bar{v}_{t,k}^{(j)}$ associated with optimal position $\hat{\pi}_{t,k}^{(j)}$. Note that this step belongs to the method called \emph{realized values} described in \cite{denault2017dynamic}.
	
	\begin{enumerate}
		\item If $t=T-\Delta t$, then 
		$$\bar{v}_{t,k}^{(j)} = u\left(\mathcal{T} \left(P_{t,k},\hat{\pi}_{t,k}^{(j)},c^{(j)}_t, \nu^{(j)}_t\right)\right).$$
		\item If $t < T-\Delta t$, then compute the realized wealth $$P_{t+\Delta t}^{(k,j)} =\mathcal{T} \left(P_{t,k},\hat{\pi}_{t,k}^{(j)},c^{(j)}_t, \nu^{(j)}_t\right),$$ and interpolate those realized wealths through the $\left\{\bar{v}_{t+\Delta t,k}^{(j)}, P_{t+\Delta t,k}\right\}_{k=1}^{n_{p}}$ pairs. Use the inverse utility interpolation method described in STEP 2b.
	\end{enumerate}
\end{enumerate}
\noindent \\
\vspace{3mm}
\underline{The Forward Step:}\\
With the choice of wealth gridpoints, the backward step becomes independent of concrete wealth developments. To obtain paths consistent  with the optimal strategy, a forward simulation is needed. Again $n_r$ new scenarios are simulated. For time $t=0,$ there is only one wealth node which coincides with the initial wealth $p$. The optimal strategy for all $n_r$ paths is then given by:
$$\hat{\pi}_{0}^{(j)} = \arg\max_{\pi \in A}\beta_{0,0}^{\intercal}\ B\left(\pi, c, \nu_0\right), \ \ j=1,\dots, n_{r}. $$
For $t\in \{\Delta t,\cdots,T-\Delta t\},$ the optimal strategy has to be interpolated between the wealth points, i.e., for $P_t^{(j)} \in [P_{t,\min},P_{t,\min}+N_{p,t}\Delta P]$, choose $\tilde{k}_{t}^{(j)}:=\lfloor(P_t^{(j)}-P_{t,\min})/\Delta P\rfloor$, such that $P_{t,\tilde{k}_{l}^{(j)}}$ is the closest grid point below $P_t^{(j)}$. Then use \eqref{eq:hatpi} with $\tilde{k}_{l}^{(j)}$ and $\tilde{k}_{l}^{(j)}+1$ (closest point above $P_t^{(j)}$) to obtain $\hat{\pi}_{t,\tilde{k}_{t}^{(j)}}^{(j)}$ and $\hat{\pi}_{t,\tilde{k}_{t}^{(j)}+1}^{(j)}$. By linear interpolation the optimal strategy is finally obtained:
$$\hat{\pi}_{t}^{(j)}=(1-\omega_t^{(j)})\hat{\pi}_{t,\tilde{k}_{l}^{(j)}}^{(j)}+\omega_t^{(j)}\hat{\pi}_{t,\tilde{k}_{t}^{(j)}+1}^{(j)},.$$
where $\omega_t^{(j)}:=(P_t^{(j)}-P_{t,\tilde{k}_{t}^{(j)}})/\Delta P$ is the weighting factor obtained from the linear interpolation.

For $P_t^{(j)} \not\in [P_{t,\min},P_{t,\min}+N_{p,t}\Delta P],$ we simply take the closest point and use \eqref{eq:hatpi} to get the optimal strategy. Note that since we do not extrapolate the wealth and the strategy outside of our gridded domains it is in particular important to choose $A$, $q_1$ and $q_2$ such that only few paths are affected.

\subsection{Parameter Specification}  
In order to learn about the effect the fluctuation of the volatility has on the solution of the retirement planning problem, parameters have to be chosen such that the differences in the solutions of the SVMRCP and the CVMRCP can be ascribed to the presence of the random volatility. Moreover, in order to obtain practically relevant results, the parameters should be consistent with observations at the financial market. For the SVM, we borrow the parameters from \cite{liu2003dynamic}, where a sound estimation procedure has been used. We summarize them in Table \ref{tab:paramsSVM}. It is well known that considerable sales of a stock lead to price decreases and increasing uncertainty that is reflected by a rising volatility. Our SVM captures this phenomenon and for that reason the correlation between the Brownian motions driving the risky asset and the volatility, respectively, is negative.  

\begin{table}[h]
	\centering	
	\begin{tabular}{llll}
		\hline\hline
		$S_0=1$, & $\nu_0=0.0169$, & $\mu=0.06$, & $r=0.02$, \\  
		 $\sigma_\nu=0.25$, &$\rho_\nu=-0.4$, & $\theta=0.0169$, & $\lambda=5$ \\
		\hline\hline
	\end{tabular}
	\caption{Input Parameters for Stochastic Volatility Model} \label{tab:paramsSVM}
\end{table}

In order to enable the aforementioned comparability between the results of the SVMRCP and the CVMRCP, certain care needs to be taken when specifying the parameters of the CVM. In the SVM, there are two market prices of risk, namely the market price of asset risk and the market price of volatility risk, while the latter is of course not present in the CVM. In order to capture and illustrate the effect of the stochastic volatility, the market prices of asset risk should coincide. This is approximately achieved by setting the volatility parameter in the CVM equal to the  long-term volatility in the SVM, so we let $\sigma_{S} = \sqrt{\theta} = 0.13.$ Furthermore, the parameters $\mu$ and $r$ as specified in Table \ref{tab:paramsSVM} are used in the CVMRCP as well. \par
The specification of the parameters of the contribution process is borrowed from \cite{chen2015target} and summarized in Table \ref{tab:paramsC}. We see that this leads to a frequently observed Sharpe ratio of 20\%. Moreover, the correlation with the risky asset is rather small. This stems from the fact that the development of a risky asset typically does not have a strong impact on the income or contribution stream of an individual and vice versa. Nevertheless, interpreting the development of a risky asset as a proxy for the overall economic situation, it is reasonable that the correlation coefficient is slightly positive.   
\begin{table}[h]
	\centering	
	\begin{tabular}{llll}
		\hline\hline
		$c=1$, & $\mu_C=0.04$, & $\sigma_C=0.1$, & $\rho_C=0.05$ \\  
		\hline\hline
	\end{tabular}
	\caption{Input Parameters for Random Contribution}\label{tab:paramsC}
\end{table}

We need to assign values to the initial wealth level, the retirement planning horizon and the relative risk aversion (RRA). This is done in Table \ref{tab:params_fund}. The initial wealth level is five times as much as the initial contribution and seems reasonable for someone starting saving for retirement. The sensitivity of the optimal strategy w.r.t. changes in the initial wealth is examined later on. Note that the choice $T=10$ means that the individual under consideration retires in ten years from now; later we investigate the sensitivity of the optimal strategy w.r.t. a prolongation of the planning horizon. A constant RRA of three is a common choice in the literature; the effect of a change is also considered later on. 
\begin{table}[h]
	\centering	
	\begin{tabular}{lll}
		\hline\hline
		$p=5$, & $T=10$, & $\gamma=3$ \\  
		\hline\hline
	\end{tabular}
	\caption{Input Parameters for the Fund}\label{tab:params_fund}
\end{table}
Finally, we need to specify the parameters of the LSMC algorithm, which is done in Table \ref{tab:params_algo}. The choice is based on the replication of solutions to well-investigated stochastic optimization problems under the constraint that the running time is reasonable. In particular, using the parameters specified in Table \ref{tab:params_algo}, the Merton problem explained below is easily solved; moreover, in \cite{kallsen2010utility} the value function of an expected utility optimization problem in a stochastic volatility model (without any contribution) similar to ours is explicitly calculated and our algorithm yields the same value of the value function (evaluated at time zero) up to several digits. This shows that the parameter choice is meaningful.  

\begin{table}[h]
	\centering	
	\begin{tabular}{llllllll}
		\hline\hline
		$N=20$, & $n_r=1000000$, & $n_{\pi}=31$, & $n_p=3$ & $\underaccent{\bar}{\pi} = -0.5$ & $\bar{\pi} = 2.5$ & $q_1 = 0.1$&$q_2=0.1$ \\  
		\hline\hline
	\end{tabular}
	\caption{Input parameters for Algorithm}\label{tab:params_algo}
\end{table}

\section{Numerical Results}
\subsection{Review}
We start the discussion of the numerical results with a brief review of the \textit{Merton Problem}, which corresponds to the optimization problem in the CVM without any contribution, and the corresponding counterpart in the SVM, which we call \textit{Heston Problem} for the sake of convenience. Dropping the contribution, the respective value functions are reduced by one dimension and the portfolio process is self-financing. \par
\begin{figure}%
    \centering
    \subfloat[Merton]{{\includegraphics[width=7.5cm]{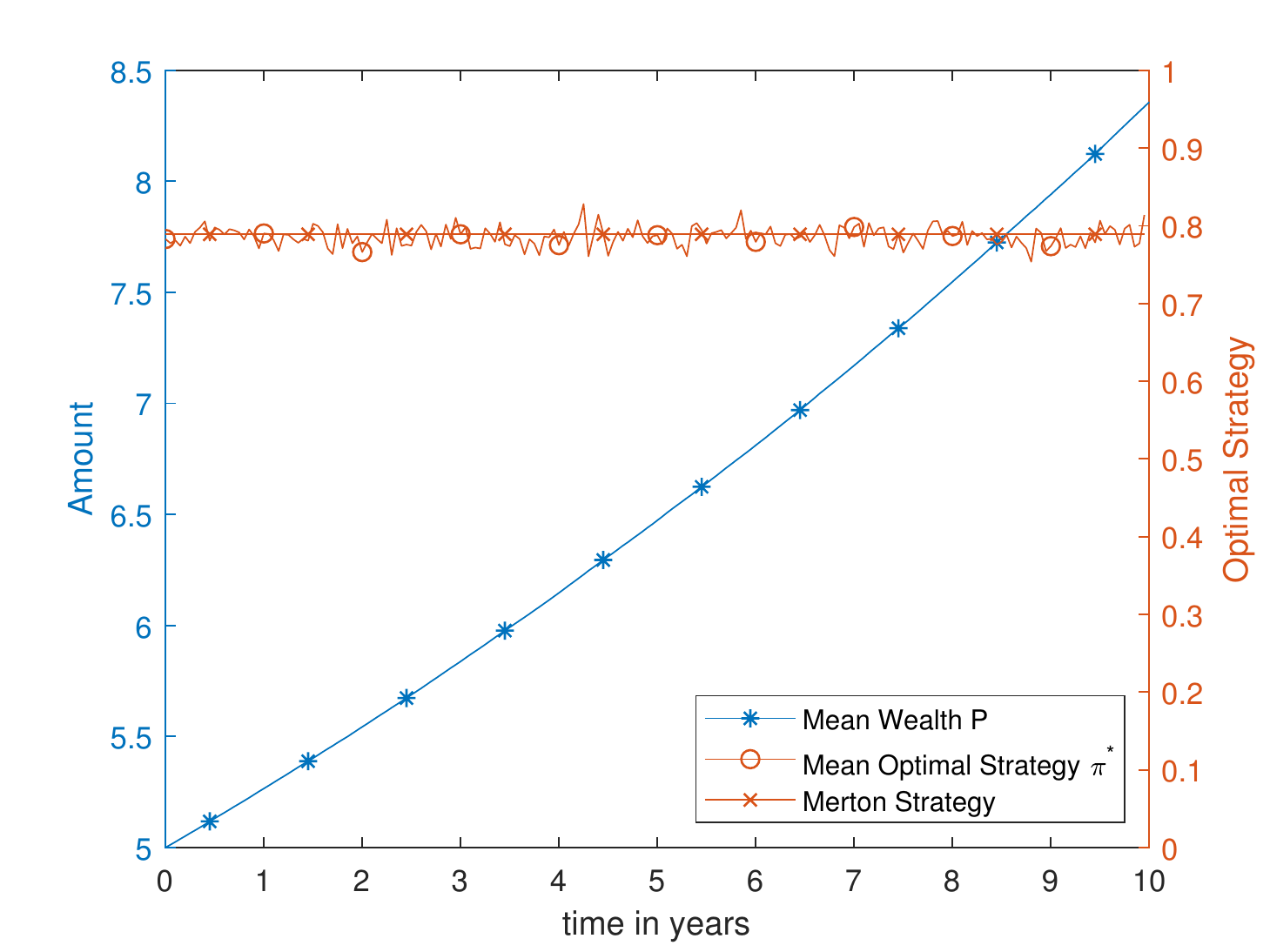} }}%
    \qquad
    \subfloat[Heston]{{\includegraphics[width=7.5cm]{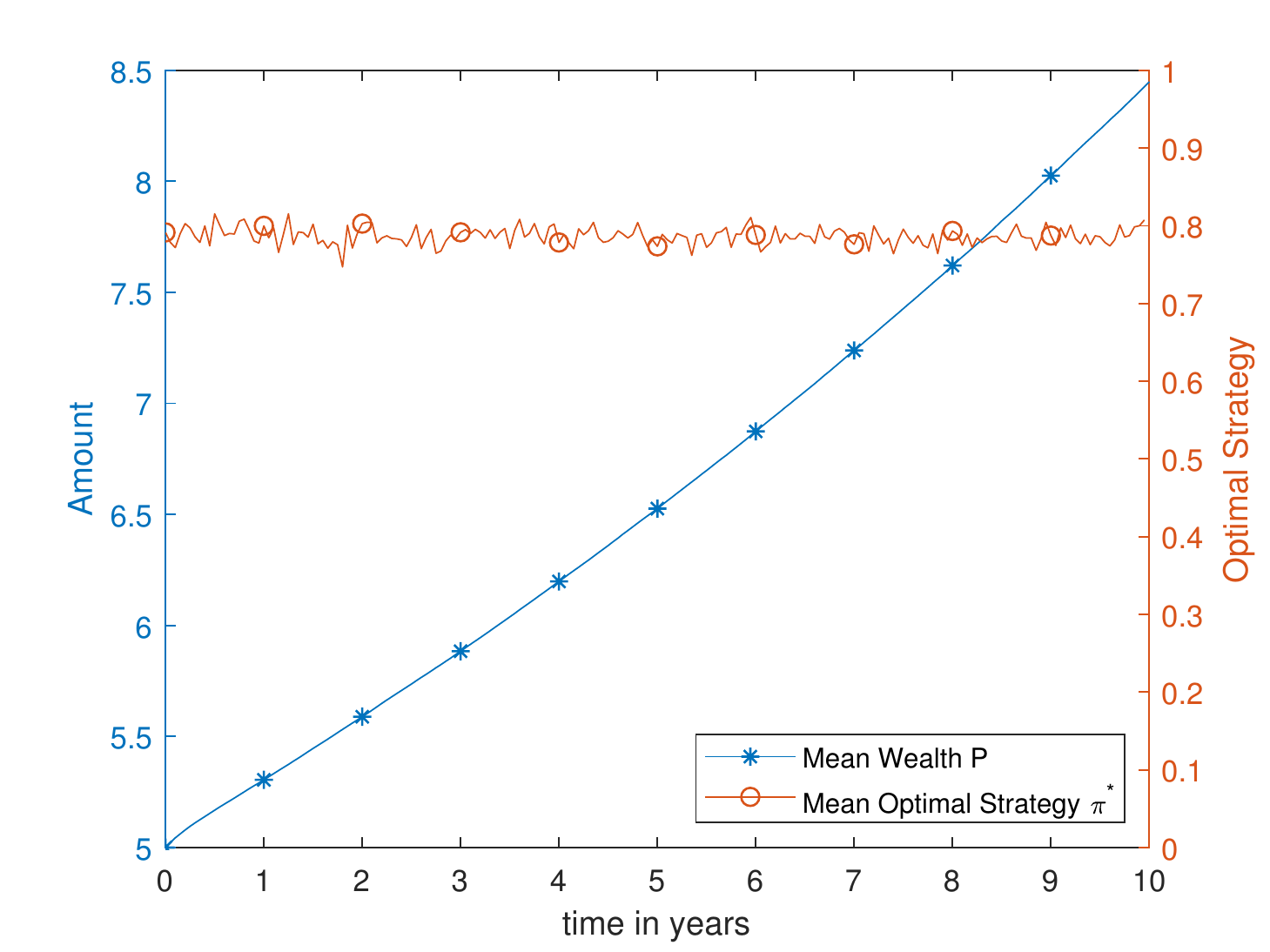} }}%
    \caption{Optimal Strategies and Wealth}%
    \label{fig:mh}%
\end{figure}

The solution to the Merton Problem, also known as Merton Ratio, is given by $(\mu-r)/\sigma^2 \gamma$ and labeled Merton Strategy in Figure \ref{fig:mh}(A). It refers to the analytical $\pi^{\star}.$ The numerically determined LSMC solution is called Mean Optimal Strategy. It corresponds to the average of optimal strategies determined by the LSMC algorithm; here and in the sequel, we only consider such averages. Obviously, the analytical and the numerical solution are relatively close to each other, thereby among others showing that the LSMC algorithm yields highly satisfactory results in this case. The optimal strategy depends on the market parameters and the risk aversion only. In particular, it is independent of time and wealth. The expected wealth induced is also depicted. The terminal value amounts to about 8.2, which corresponds to an average increase of 5\% per year. Thus, on average the Merton portfolio outperforms an investment in the riskless asset solely. Similar remarks apply in the case of the Heston Problem, which is graphically depicted in Figure \ref{fig:mh}(B). The strategy seems also independent of time.  We see that it is fluctuating around the optimal fraction to be invested in the Merton Problem with partly larger spikes than in the Merton numerical solution. However, as the volatility in the CVM is chosen equal to the long-term volatility in the CIR-process, the mean-reversion property of the latter causes the solution in the Heston problem to be relatively close to the Merton Ratio. The terminal expected wealth nearly coincides with the one generated by the Merton portfolio in the CVM.

\subsection{The CVMRCP and the SVMRCP}
\begin{figure}%
    \centering
    \subfloat[CVMRCP]{{\includegraphics[width=7.5cm]{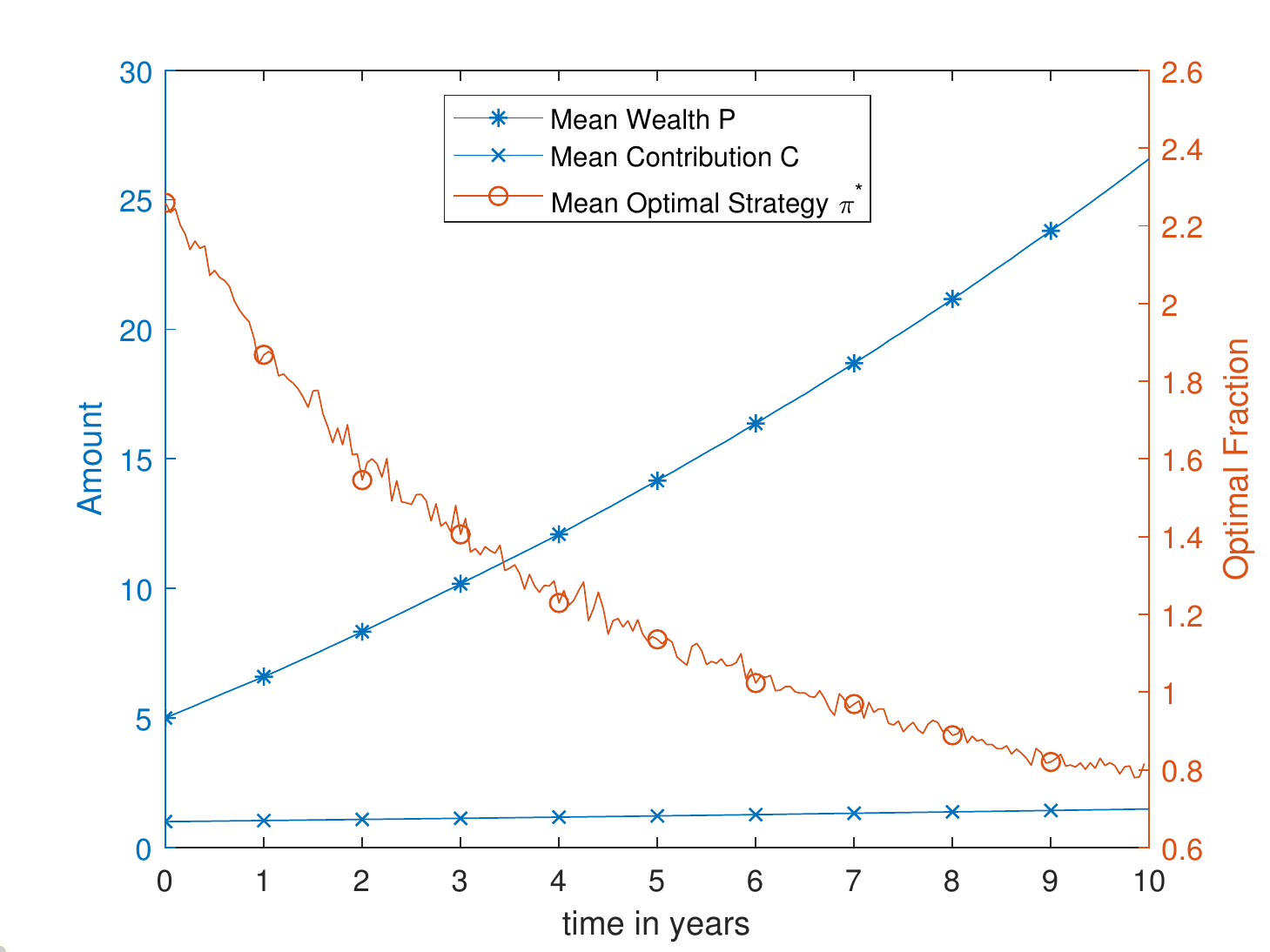} }}%
    \qquad
    \subfloat[SVMRCP]{{\includegraphics[width=7.5cm]{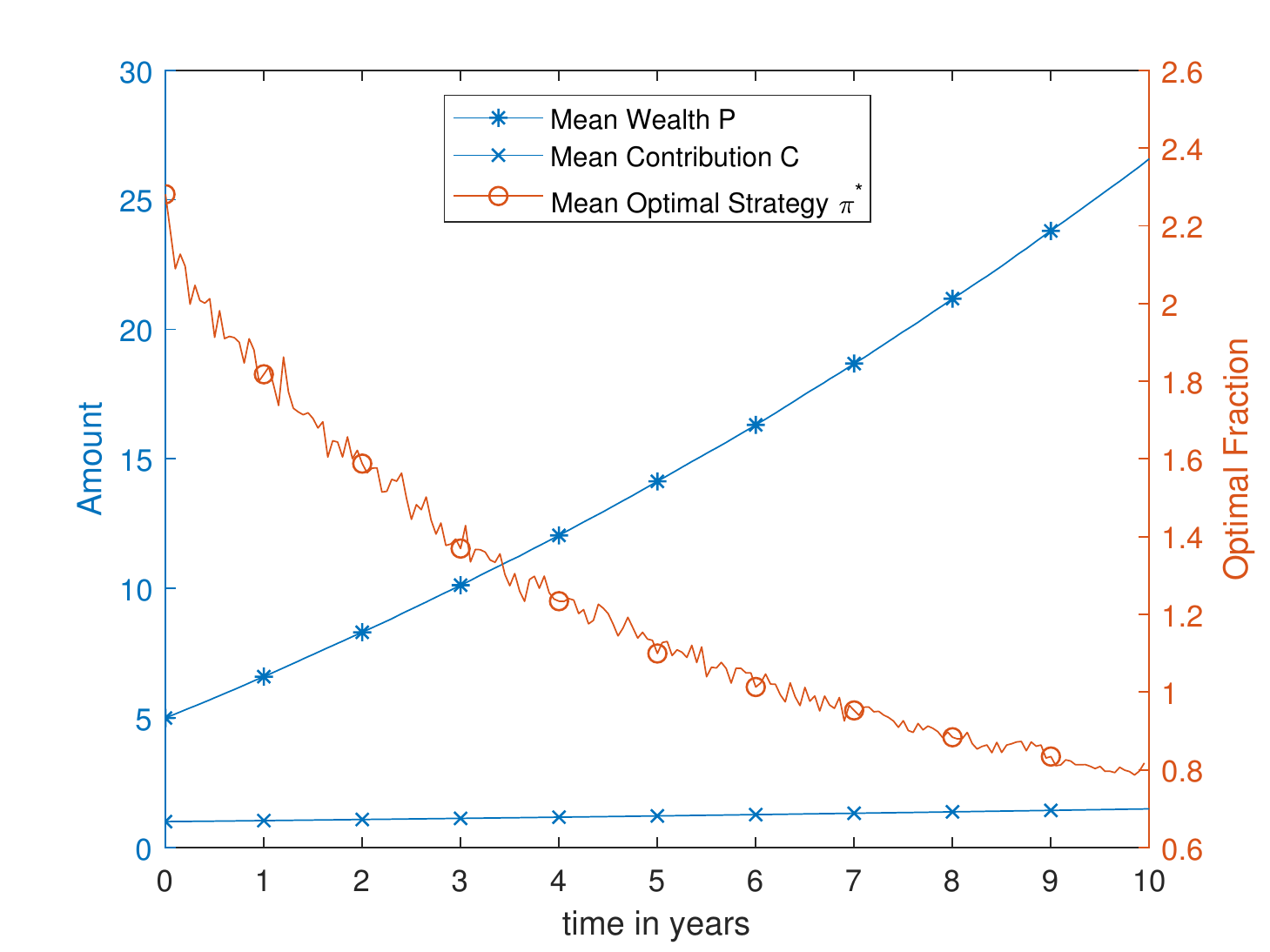} }}%
    \caption{Optimal Strategies and Wealth with Random Contribution}%
    \label{fig:opt_strat}%
\end{figure}

In this subsection we present the solutions to the CVMRCP and the SVMRCP, respectively, using the parameters specified in Section 3.2 above. The numerical results are displayed in Figure \ref{fig:opt_strat}. In the left panel we see the optimal fraction to be invested in equity in the CVM, the wealth induced and the random contribution (all quantities are again to be understood as averages). We note that the fraction of wealth invested in the risky asset is - up to minor local movements - decreasing over time. The decay is obviously larger at the beginning than towards maturity. For example, in the first five years we observe a reduction of more than 100 percentage points while in the second five years the decrease amounts to about 20 percentage points. Moreover, the spikes are more visible in the first period mentioned. Hence, the investment behaviour becomes less risky as time progresses. Over ten years, the optimal expected wealth increases by roughly 380 \%, which corresponds to an average annual growth (return of currently wealth plus contribution from income) of more than 16 \%. Note that the annual growth in percentage in general is higher in the early phase of the investment because the financial returns are supplemented with contributions from income which take a relative large share.
In the right panel of Figure \ref{fig:opt_strat} the solution to the SVMRCP is displayed. Overall, we can see several similarities to the constant volatility case of the left panel: the optimal trading strategy is decreasing over time and the decay is stronger in the first half of the savings period than in the second half. The development of the expected terminal wealth and therefore the part that can be ascribed to the investment in the risky asset is nearly the same. Measuring the difference of wealth invested in the risky asset, only negligible differences can be observed.\par
In order to gain further insight into the effect the inclusion of the stochastic volatility has on the optimal wealth process, Table \ref{table_T} contains Mean, Variance and Certainty Equivalent (CE)\footnote{The non-random amount yielding the same level of expected utility as the optimal terminal wealth of the portfolio: $\text{CE} = u^{-1}\left(\mathbb{E}_{t,p,x,c}[P_{T}^{\pi^{\star}}]\right).$ If not explicitly  stated otherwise, we refer to the CE seen from time point $t=0.$} of the optimal terminal wealth; they are to be understood as empirical quantities. Likewise, we investigate the effect of a change of the correlation coefficient between the Brownian motions driving the risky asset and the stochastic variance. Obviously, the expected values in the CVM and the SVM for all levels of correlation considered nearly coincide. Regarding the variation, we see that the variance of the terminal wealth in the SVM is increasing as $\rho_{\nu}$ is decreasing - this effect can be explained as follows: an increasing correlation coefficient $\rho_{\nu}$ leads to an increasing variance of $S_{t}$ for any $t \in (0,T].$ A higher variance is identified with a higher riskiness of the investment  and consequently leads to a reduction in the fraction of wealth invested in $S$. At first sight it is not clear which effect dominates. Indeed, considering the original Merton problem (set the contribution to zero in the constant volatility case) with optimal strategy $(\mu-r)/(\sigma_S^2\gamma)$ and optimal portfolio process
$$dP^{\text{Merton}}_t= P^{\text{Merton}}_t\left(r +\frac{(\mu-r)^2}{\gamma \sigma_S^2}\right) dt+P^{\text{Merton}}_t\frac{(\mu-r)}{\gamma \sigma_S}dW_t^{(1)},$$
in the denominator of the resulting stochastic integral there is $\sigma_{S}$ as multiplicative factor. That is, an increase of the volatility of the stock leads to a decreasing variance of the portfolio process.
Our results show that the same logic applies in the SVMRCP. Thus, a low correlation, which leads to a lower variance of the risky asset, induces a relatively higher investment. Table \ref{table_T} then shows that the effect of the variance of $S$ on the optimal trading strategy is so strong that the variance of the optimal terminal wealth is finally the highest for the originally least risky case of $\rho_{\nu} = -0.9,$ and that it is the lowest for the actually riskiest case in which $\rho_{\nu} = 0.9.$ Hence, the optimal trading strategy counteracts the impact of the correlation coefficient on the variance of the final wealth. The same remarks apply to the behaviour of the CE considered in dependence on $\rho_{\nu}.$ Furthermore, we see that the variance of the optimal wealth in the CVM is higher than in the SVM for all levels of correlation considered. Using the log-normal return distribution, it is straightforward to calculate that the variance of the stock in the CVM is achieved in the SVM for a correlation around 0.2. The values of the variance shown in Table \ref{table_T} clearly imply that for this value of the correlation the variance in the SVM is smaller than in the CVM. A possible explanation for the higher variance in the CVM is that an investor with a risk aversion of $\gamma =3$ perceives the CVM as relatively safe, leading to a higher fraction of wealth invested in the risky asset. This higher fraction of wealth then counteracts the safety of a constant volatility environment. 
\begin{table}
	\centering
	\begin{tabular}{l c| c c c c c}
		&\text{CVM}& & & \text{SVM} \\ \hline \hline
		$\rho_{\nu}$ 						& \%		& 0.9 	& 0.4  & 0.0 & -0.4&-0.9\\ \hline
		$\mathbb{E}[P^{\pi^{\star}}_{T}]$ 	& 26.51 	&26.56 	& 26.56&26.56&26.57&26.53  \\ \hline
		$\text{Var}[P^{\pi^{\star}}_{T}]$ 	&83.87 		&62.59 	& 67.04&70.65&74.19&77.88 \\ \hline 	 
		$\text{CE}$ 						& 22.27 	&23.22 	& 23.08&22.97&22.86&22.71 \\ \hline \hline
	\end{tabular}
	\caption{Expectation, Variance and Certainty Equivalent (CE) in the two models, $\gamma=3$}
	\label{table_T}
\end{table}
	\par
Reasoning from the similarities of the optimal strategies in Figure \ref{fig:opt_strat}(A) and Figure \ref{fig:opt_strat}(B) and the almost coincidence of the expected values in the two models in Table \ref{table_T}, the question arises whether there is a strong effect when applying for instance the solution of the CVMRCP in the SVMRCP. This corresponds to the situation that an investor presumes the constant volatility setting and determines her optimal strategy, while the instantaneous variance actually develops according to the CIR-process. This yields an expected value of the terminal wealth of 26.51, a variance of 81.71 and the CE then amounts to 22.20. Hence, drawing comparisons with the  column showing the case $\rho_{\nu} = - 0.4$ of Table \ref{table_T}, the erroneously followed strategy yields a marginally lower expected payoff and CE, while the variance increases by roughly 10.5 \%. Hence, although the strategies show a very similar qualitative behaviour, ignoring the presence of random volatility leads to a higher risk exposure.  \par

\begin{figure}%
	\centering
	\subfloat[SVM]{{\includegraphics[width=5.5cm]{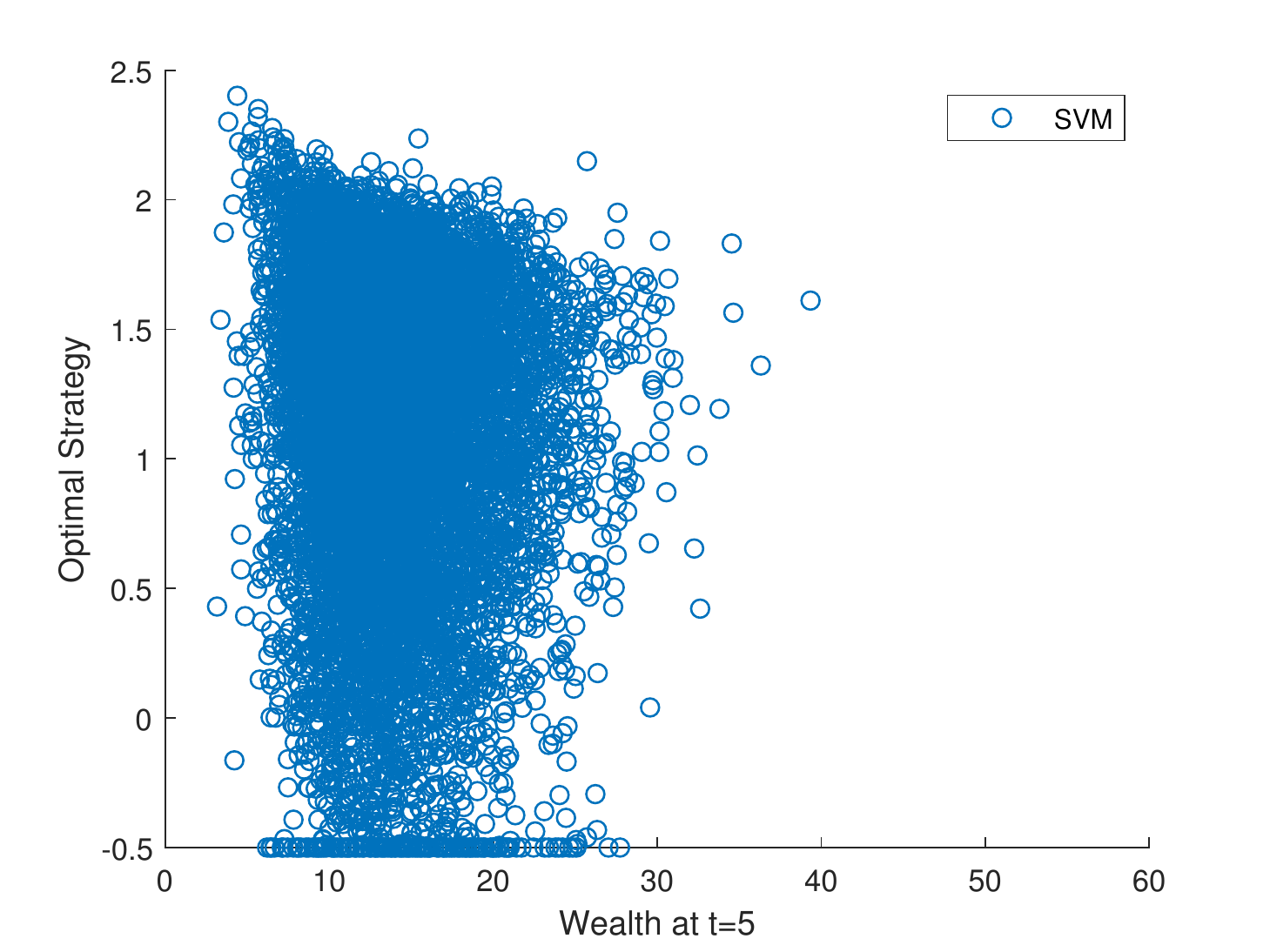} }}%
	\subfloat[SVM $v_t \in {[0.01,0.02]}$ ]{{\includegraphics[width=5.5cm]{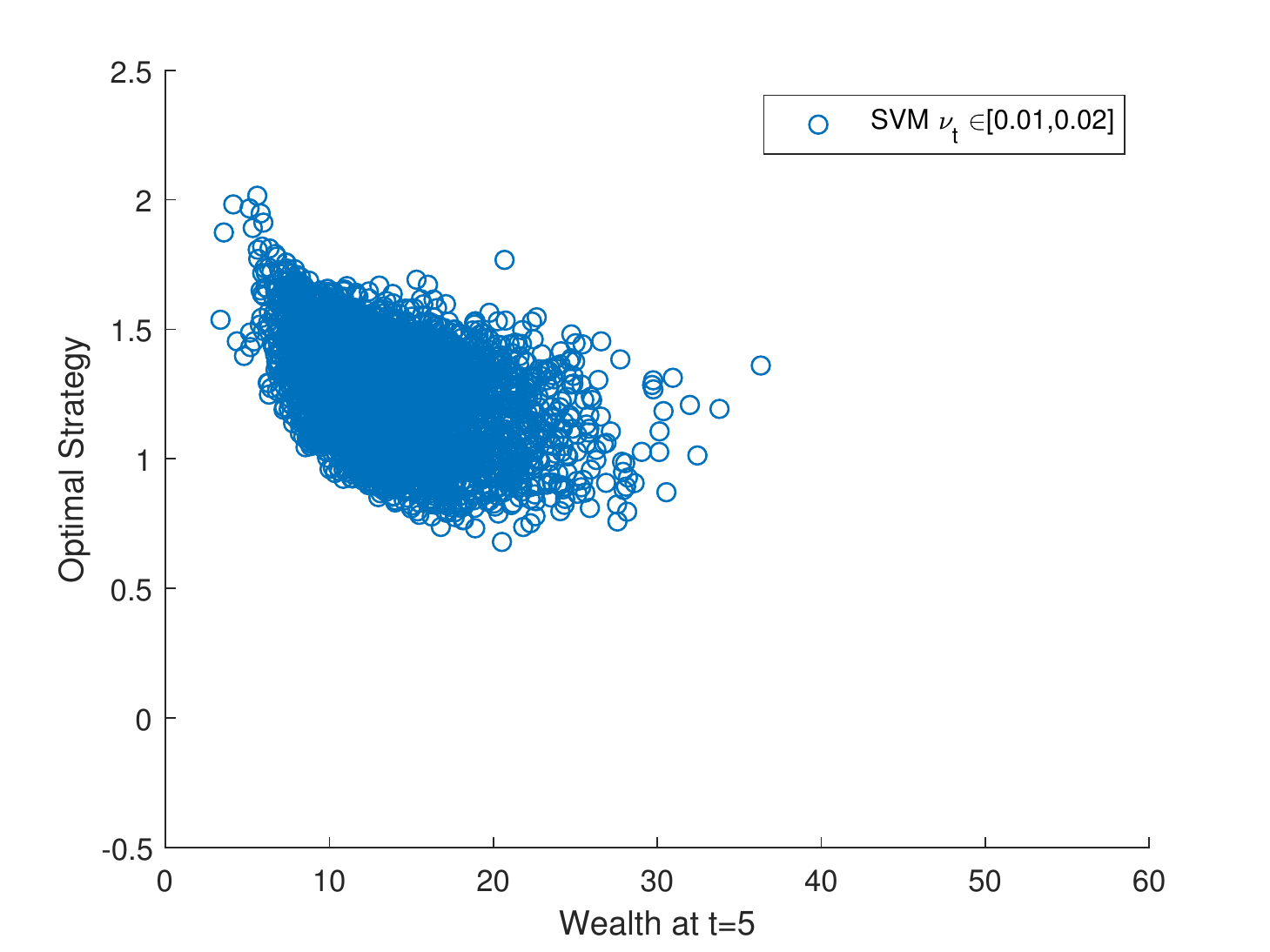} }}%
	\subfloat[CVM]{{\includegraphics[width=5.5cm]{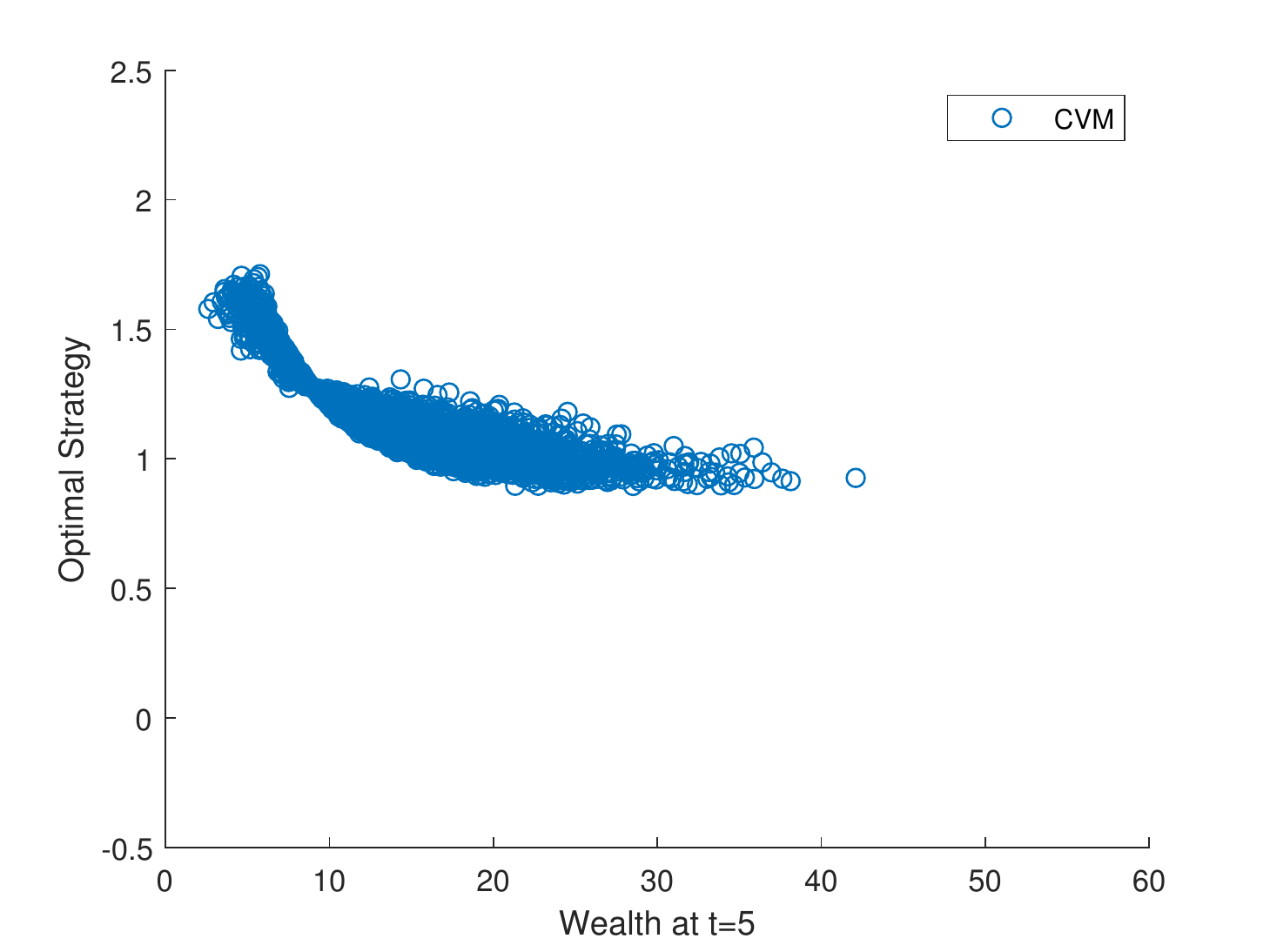} }}%
	\caption{Scatter Plot Optimal Strategies in Stochastic Volatility Model and Constant Volatility Model}%
	\label{fig:scatter}%
\end{figure}

So far, our analysis has focused on the average of optimal strategies over one million paths. In order to further learn about the impact of stochastic volatility, we now broaden our perspective. In Figure \ref{fig:scatter}, the time point is fixed to five years after the start of the planning horizon and the x-axis shows different possible wealth levels. For each path, after observing the corresponding wealth level at time $t=5,$ we plot the respective optimal fraction to be invested. The left panel thereby corresponds to the SVM and the right panel to the CVM. It is obvious that for the same wealth levels the amplitude of the optimal fraction of wealth to be invested in the risky asset is much larger in the SVM than in the CVM. Thus, the stochastic volatility explains a considerably larger range of possible investment fractions. The middle panel resembles the case that the stochastic volatility lies between 10\% and 20\%, i.e., the stochastic volatility is in a narrow neighborhood of the constant one. For these values of the volatility, we see that the strategy is closer to the optimal one determined in the CVM. It also shows a similar dependence on the wealth level. Hence, the large range of strategies explained by the SVM yields average results that nearly coincide with the CVM.

The omission of the random contribution reduces the CVMRCP to the Merton Problem and the SVMRCP to the Heston Problem. We see that the contribution's inclusion in either case induces an investment strategy that is decreasing over time with a considerably higher fraction of wealth to be invested in equity at the beginning. Put differently, the investment behaviour is initially much more risk-affine. This is due to the fact that the investor knows fully well that potential losses might be offset by the continuous (future) contribution stream which makes her accept a higher level of riskiness. As the time point of retirement approaches, the future contribution stream becomes relatively low and the risk exposure is diminished through a reduction of the fraction of wealth invested in the risky asset.  \par

\subsection{Impact of Initial Wealth and Initial Contribution}
\begin{figure}%
    \centering
    \subfloat[$P_0\in\{2.5,5,10\}, C_{0} = 1$]{{\includegraphics[width=7.5cm]{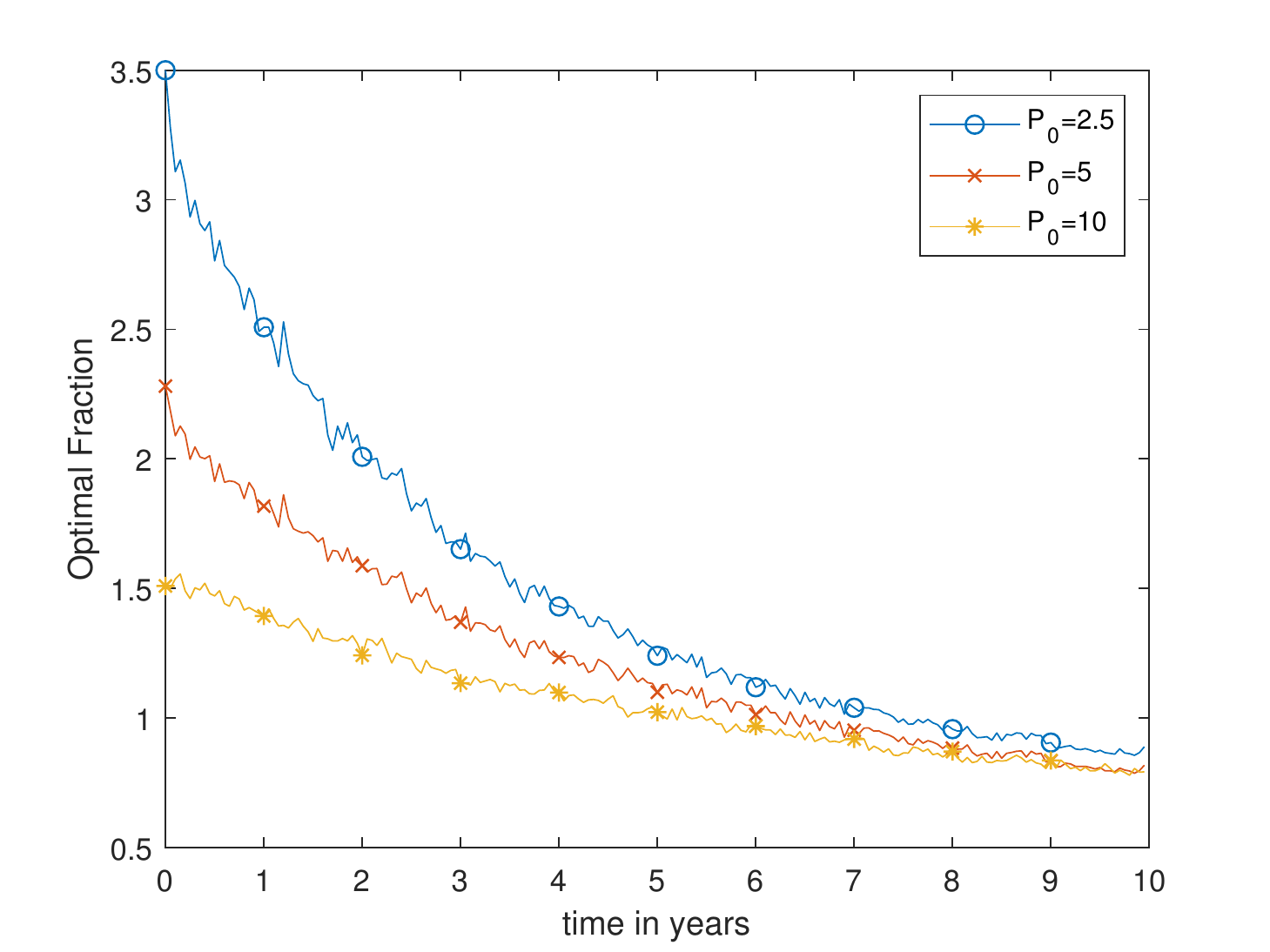} }}%
    \qquad
    \subfloat[$C_{0} \in \{0.5,1,2\}, P_0 = 5$]{{\includegraphics[width=7.5cm]{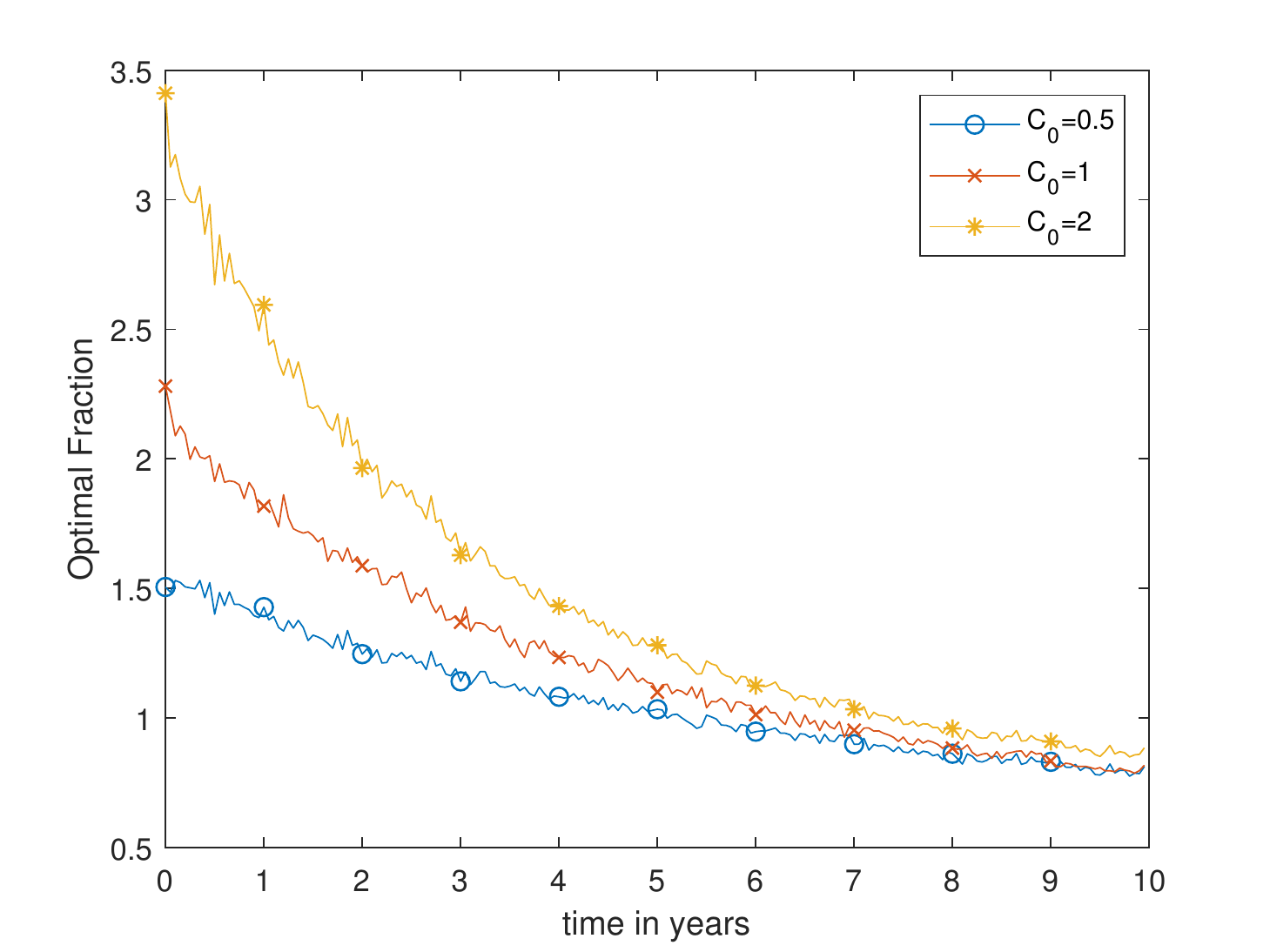} }}%
    \caption{Impact of Initial Wealth (left) and Initial Contribution (right), Stochastic Volatility Model}%
    \label{fig:p_and_c}%
\end{figure}

We next investigate the effect the initial wealth and contribution levels have on the optimal strategies. Since the observations in the CVM and the SVM were very similar so far, we only consider the SVM in this subsection and remark that analogue results are found for the CVM. The sensitivity analysis with respect to the initial wealth and the human capital (the initial value of the total future contributions) is important, as various combinations of the initial wealth and human capital can represent pension beneficiaries with different ages. For instance, a young beneficiary typically has a lower initial wealth, but a higher human capital. On the contrary, an older beneficiary probably has accumulated quite some wealth so far and will start with a higher initial wealth level, but carry less human capital, as she will work shorter than the younger one. These analyses help us understand the investment behavior of pension beneficiaries with different ages. \\
In the left panel of Figure \ref{fig:p_and_c} the behavior of the optimal investment strategy for different initial wealth levels is depicted. Hereby, we have assumed that the initial contribution $C_0$ stays unchanged, i.e., the initial value of total future contributions (human capital) does not change. Reasoning from the graphic, this implies that the pension beneficiary who owns less initial wealth $P_0$ (upper curve), but the same level of human capital, invests a higher fraction of wealth in the risky asset.
A higher initial wealth goes along with a higher certainty of ending up with a large final wealth. Consequently, the necessity to take on risk to generate money for retirement is reduced and an investor can follow a more conservative strategy. As time progresses, the strategies starting with different wealth levels become closer to each other. This is a consequence of using a utility function with CRRA which is common in economics as, we recall, the latter implies that the long-term behaviour of economic agents (in this case the retirement planner) is independent of wealth.  \par
In the right panel of Figure \ref{fig:p_and_c} the optimal strategies for different initial levels of contribution are displayed. 
Hereby, we hold the initial wealth $P_0$ fixed and vary the initial contribution $C_0$. The higher $C_0$, the more human capital the pension beneficiary owns. An increase of $C_0$, i.e. an increase in the human capital, fosters investment in the risky asset (upper curve). To some level, it can be seen as the pension beneficiaries tend to borrow the future income to invest more in the risky asset.
In both panels, we observe that the higher the ratio $P_0/C_0$ (realized by a lower initial contribution level $C_0$ or a higher initial wealth level $P_0$), the lower the resulting optimal fraction invested in the risky asset.

\subsection{Effect of Risk Aversion} 
\begin{figure}[h]
	\centering
	\includegraphics[width=9cm]{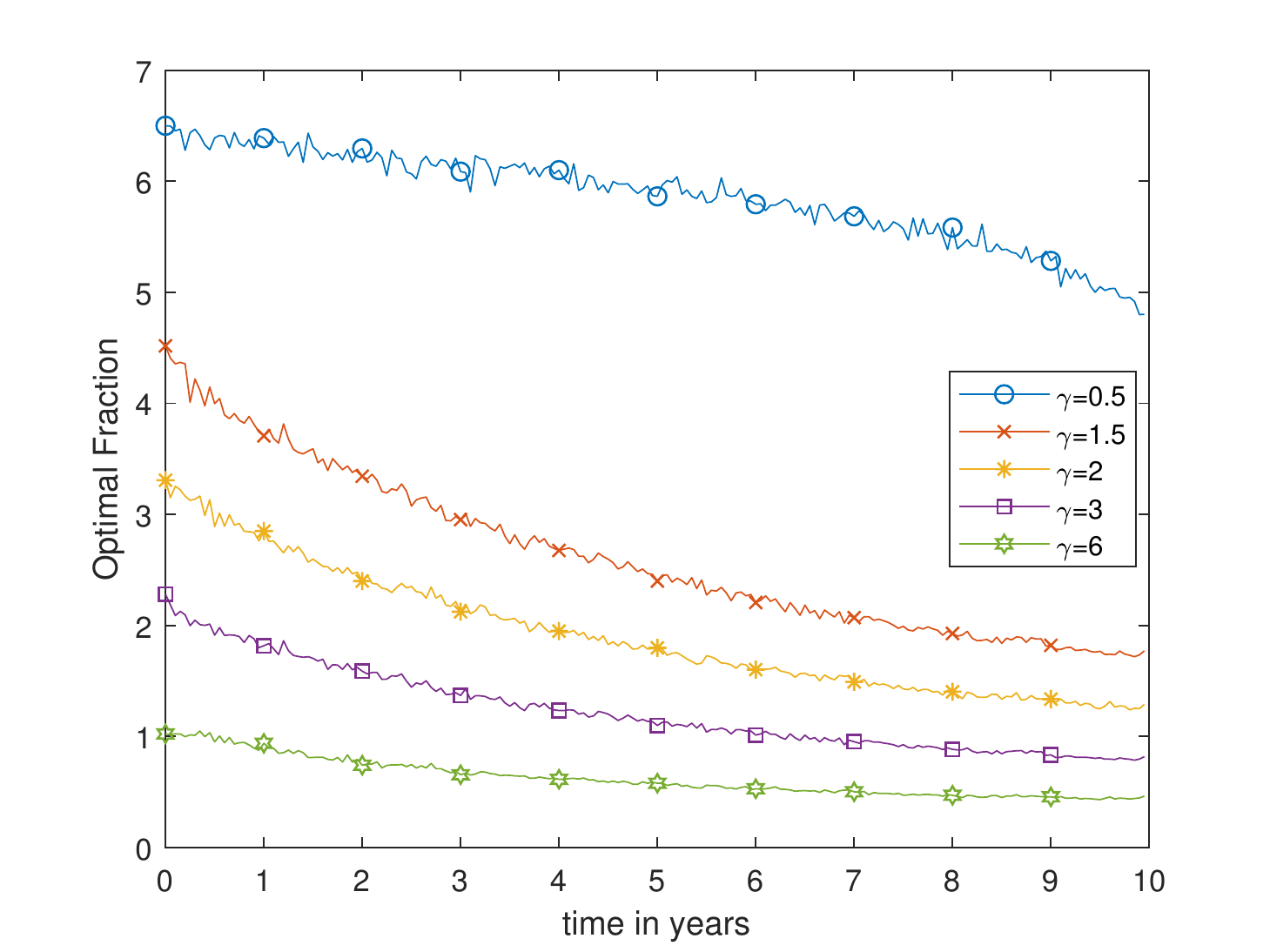}
	\caption{Optimal Trading Strategy for Different Levels of Relative Risk Aversion, Stochastic Volatility Model}
	\label{fig:SV_Gamma}
\end{figure}

A close look at the Merton Ratio reveals that the fractional amount to be invested in equity is decreasing as the RRA is increasing. In this section we investigate whether this effect can also be observed if a random contribution is present and when stochastic volatility is included. In Figure \ref{fig:SV_Gamma} we depict the optimal investment strategy over time for different levels of risk aversion. With increasing RRA, the investment in the risky asset is reduced. Moreover, we observe that as $\gamma$ increases, the shape of the optimal strategy changes from the glidepath type to a horizontal line as in the Heston solution shown in Figure \ref{fig:mh}(B). This means that the impact of the contribution causing the shape of the glidepath is diminished when the RRA is becoming large. \par
 In Table \ref{tab:gammaSens} the CE, Mean and Variance of the optimal terminal wealth is depicted for different levels of risk aversion in the SVM. We observe that the CE, Mean and Variance are decreasing when the RRA is rising. This is a consequence of the reduction of the fraction of wealth to be invested in equity. Looking again at the lines in Figure \ref{fig:SV_Gamma}, it is apparent that the curve resembling the strategy for $\gamma = 0.5$ is considerably higher than all the others, and the marginal difference between this one and the curve showing the strategy for $\gamma = 1.5$ is the largest. This observation is also numerically depicted in Table \ref{tab:gammaSens} as the marginal gap between CE, Mean and Variance is the biggest when increasing the RRA from $0.5$ to $1.5$. Thus, the effect on the strategy when changing from a risk-affine investor (corresponding to the case $\gamma < 1$) to a risk-averse investor, say a value of $\gamma = 1.5$, is larger than the effect of becoming even more risk-averse. 

\begin{table}[h]
	\centering	
	\begin{tabular}{c|ccc}
		 RRA & CE & $\mathbb{E}[P_{T}^{\pi^{\star}}]$  & $\text{Var}[P_{T}^{\pi^{\star}}]$\\ 
		\hline\hline
		 $0.5$ &$53.76$ & $103.11$ & $173305.67$\\
		 $1.5$ &$28.33$ & $38.10$ & $763.15$\\
		 $2$ &$25.46$ & $31.73$ & $259.57$\\
		 $3$ &$22.86$ & $26.578$ & $74.19$ \\
		 $6$ &$19.70$ & $22.94$ & $33.21$\\ \hline \hline
	\end{tabular}
	\caption{Comparison of Different Levels of Relative Risk Aversion, Stochastic Volatility Model}\label{tab:gammaSens}
\end{table}

\subsection{Impact of the Contribution's Parameters}
\begin{figure}[h]
	\centering
	\includegraphics[width=9cm]{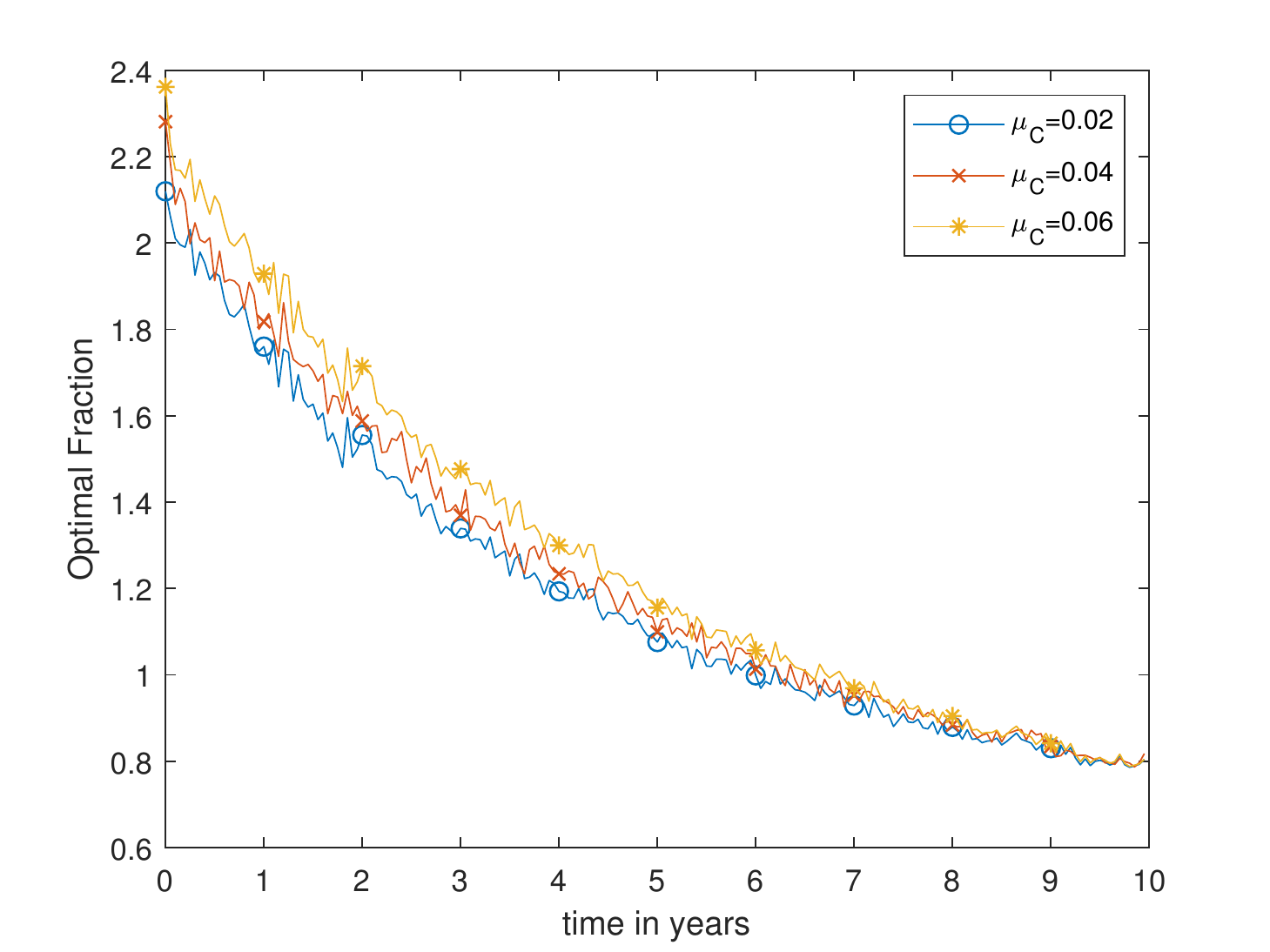}
	\caption{Optimal Trading Strategy for Different Levels of $\mu_C$, Stochastic Volatility Model}
	\label{fig:CV_Mu}
\end{figure}

We recall the previous result that a low wealth-to-contribution ratio leads to a rather risky investment strategy. In this subsection we further investigate the marginal impact of the contribution on the optimal share to be invested in equity. To this end, we vary the parameters in the contribution process.
First, we let the contribution's drift $\mu_{C}$ change, which only has a relatively small impact on the optimal strategy, see Figure \ref{fig:CV_Mu}. Overall we find that a higher drift makes the agent willing to invest more riskily initially. This result is consistent with our previous findings as a higher drift leads in general to a higher level of contribution, which in turn decreases the aforementioned wealth-to-contribution ratio. However, this effect gradually declines over time and the optimal strategies depicted in Figure \ref{fig:CV_Mu} show a nearly identical behaviour at maturity. This is a consequence of the fact that the wealth is dominating the wealth-to-contribution ratio as time progresses and therefore the effect of a marginally higher contribution vanishes. In Table \ref{tab:muSens} an analysis of the final wealth's CE, Mean and Variance is performed for different levels of $\mu_{C}$ presuming a deterministic contribution stream (upper panel) and the stochastic contribution process (middle and lower panel); as the CVM and the SVM show similar results in this case, we restrict the analysis to the latter. Note that $\mu_{C} = \sigma_{C} \equiv 0$ means that the contribution is kept at the initial level $C_{0}$ and for every period of length $\Delta,$ the amount $C_{0} \cdot \Delta $ is contributed to the pension fund. For the situation that the contribution is deterministic, an increase of the contribution's drift naturally leads to a higher CE, Mean and Variance. Looking at the lower panel of Table \ref{tab:muSens}, we keep $\sigma_{C}$ at a value of 0.1 and vary $\mu_{C}.$ The same effects as in the previously discussed deterministic contribution case are observed. \par
Similar to the drift, we analyze the impact of the volatility of the income process $\sigma_C$. Comparing the deterministic with the stochastic income process (both for the value $\mu_{C} = 0.04$), we see that the deterministic stream yields a higher CE, a negligibly higher expected wealth and a slightly lower variance. Thus, the presence of randomness in the contribution process only adds slightly to the riskiness of the investment procedure.  \par
\begin{figure}[h]
	\centering
	\includegraphics[width=9cm]{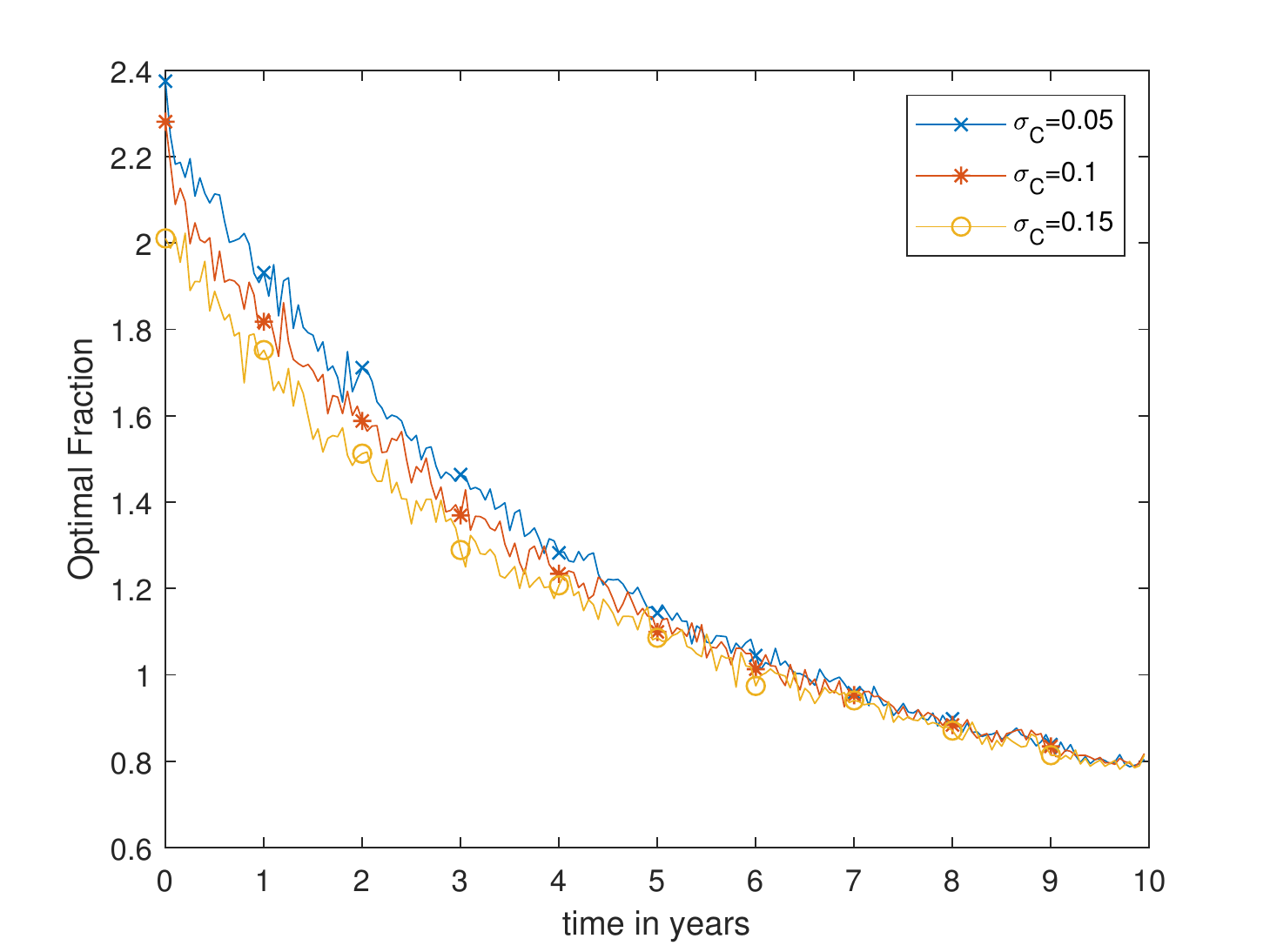}
	\caption{Optimal Trading Strategy for Different Levels of $\sigma_C$, Stochastic Volatility Model}
	\label{fig:SV_Sigma}
\end{figure}

 In the lower panel of Table \ref{tab:muSens}, we vary the value of $\sigma_{C}$ while keeping $\mu_{C}$ constant. The CE and the expected terminal wealth are not very sensitive to changes in the volatility of the contribution process. However, a higher volatility of the contribution process adds some variation to the final wealth. Figure \ref{fig:SV_Sigma} contains a graphical illustration. Overall we see that with increasing volatility the agent invests slightly less risky in the early stages of the contract while the strategies for different values of $\sigma_{C}$ show a very similar behaviour as maturity approaches. However, the overall effect is small. Our results show that the main factor of the contribution process effecting the optimal strategy is the initial level.  
\begin{table}[h]
	\centering	
	\begin{tabular}{cc|ccc}
		$\sigma_C$ & $\mu_C$ & CE & $\mathbb{E}[P_{T}^{\pi^{\star}}]$  & $\text{Var}[P_{T}^{\pi^{\star}}]$ \\
	    \hline\hline
		 $0.0$ & $0.0$ &$20.47$ & $23.35$ & $51.26$  \\
		 $0.0$ & $0.04$ & $23.48$ & $26.89$ & $67.73$\\
		\hline\hline
		 $0.1$ & $0.02$ &$21.49$ & $24.88$ & $64.85$  \\ 
		 $0.1$ & $0.04$ &$22.86$ & $26.578$ & $74.19$ \\ 
		 $0.1$ & $0.06$ &$24.59$ & $28.53$ & $85.41$ \\   
		\hline\hline
		 $0.05$ & $0.04$ & $23.41$ & $26.79$ & $69.49$ \\  
		 $0.10$ & $0.04$ &$22.86$ & $26.578$ & $74.19$\\  
		 $0.15$ &$0.04$&$22.23$ & $26.34$ & $83.66$ \\
	\end{tabular}
	\caption{Sensitivity to Varying Parameters of the Contribution Process, Stochastic Volatility Model}\label{tab:muSens} 
\end{table}

\subsection{Prolongation of the Planning Horizon}

\begin{figure}%
    \centering
    \subfloat[CVMRCP]{{\includegraphics[width=7.5cm]{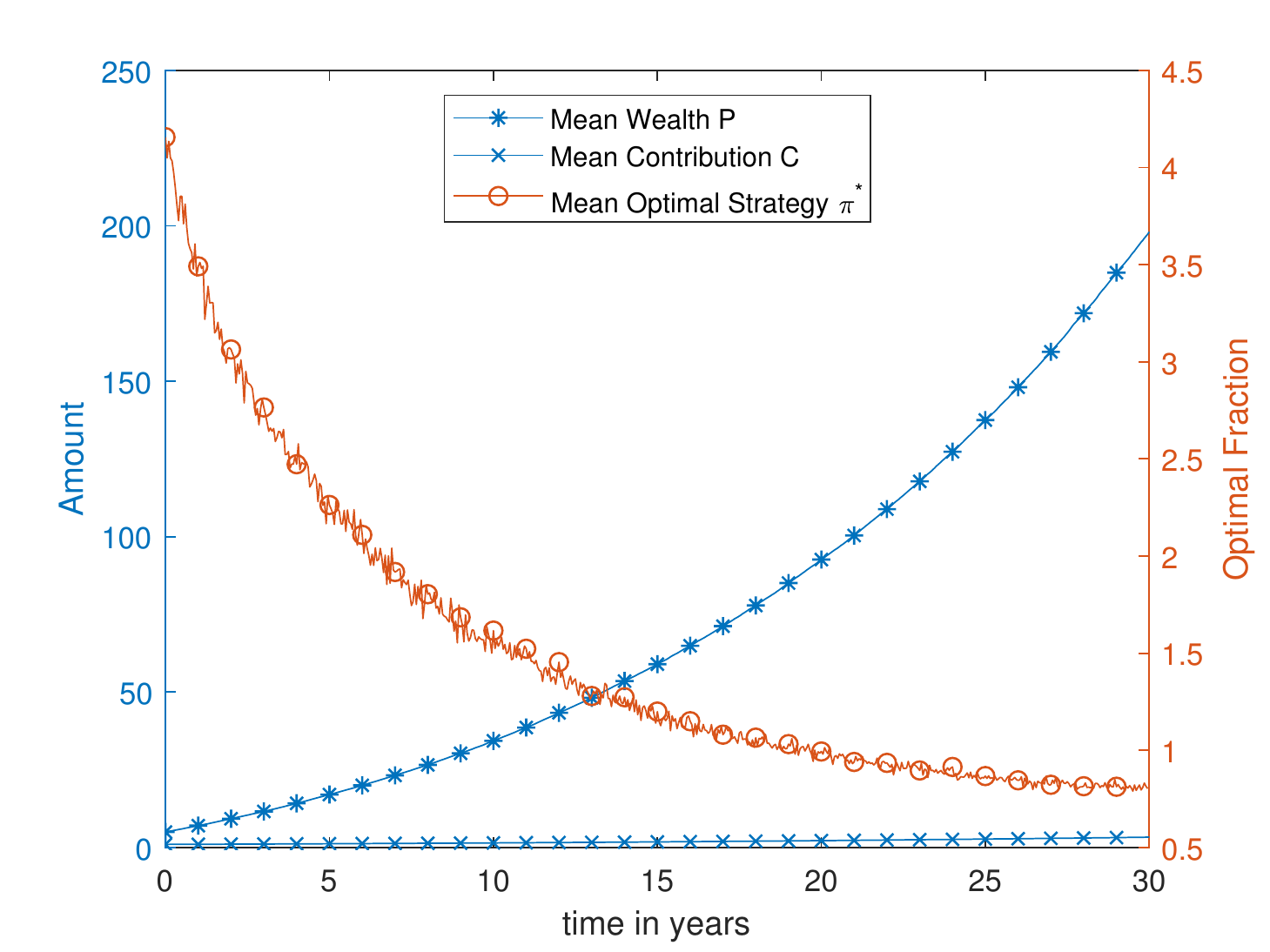} }}%
    \qquad
    \subfloat[SVMRCP]{{\includegraphics[width=7.5cm]{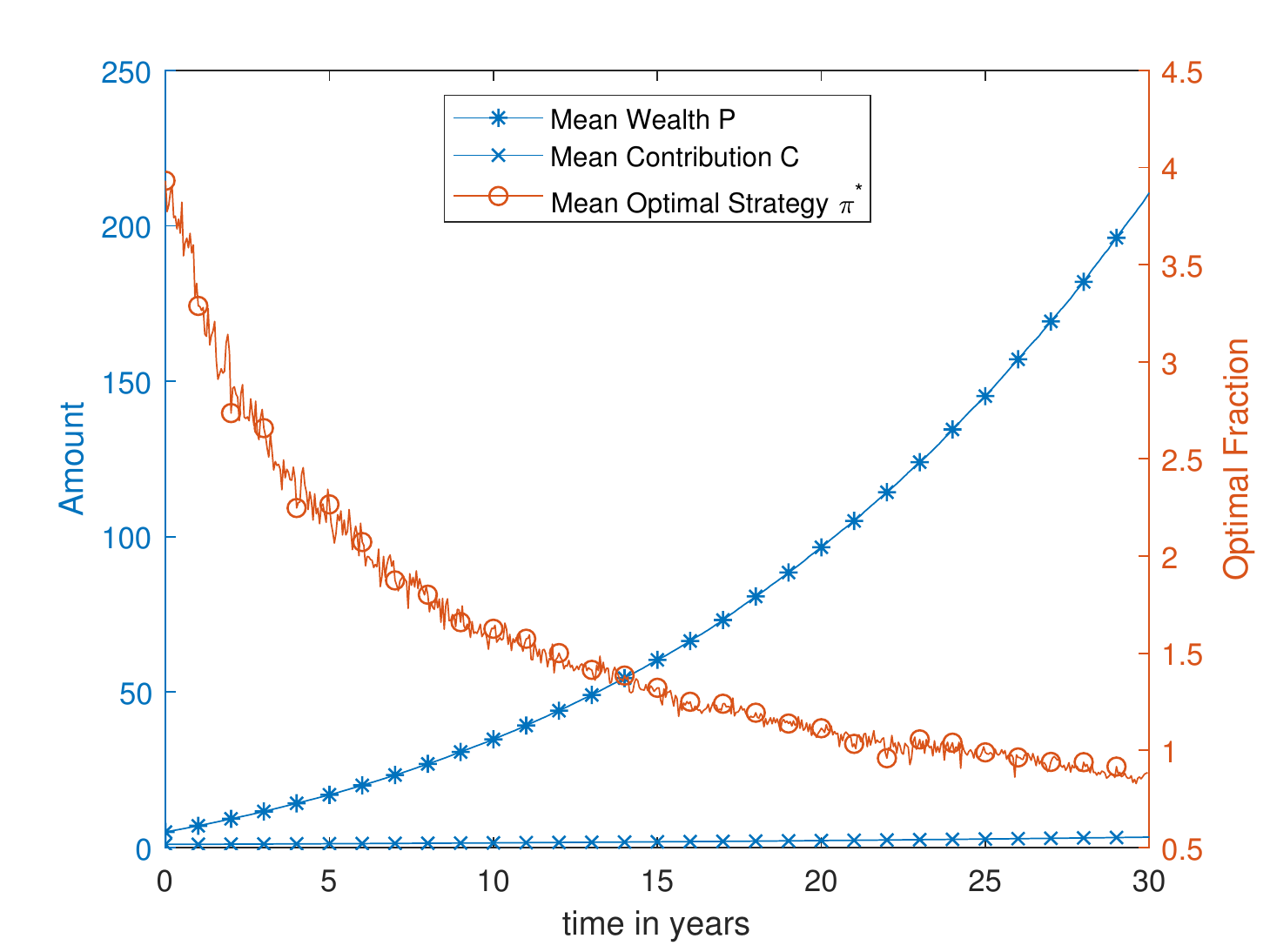} }}%
    \caption{Optimal Strategies and Wealth with Random Contribution, $T=30$}%
    \label{fig:prolong}%
\end{figure}

So far we have considered a planning horizon of $T=10$ years. This usually corresponds to a retirement planner in the second half of his fifties. As more and more people start saving for their pension when they are younger, we consider the case $T=30$ in this subsection. This comes along with a significant increase of the computational effort stemming from the increased number of wealth nodes. For longer time horizons such as $T=30,$ we therefore use a slightly modified version for the calculation of the step-size which is time-dependent and recalculated on an annual basis (using the notation introduced in the Prerequisites of the LSMC algorithm above):
	$$\Delta P_t:=\frac{P_{\Delta t+ \lfloor t\rfloor,\max}-P_{\Delta t+\lfloor t\rfloor,\min}}{n_p +\lfloor t\rfloor}.$$ 
Figure \ref{fig:prolong} shows the corresponding optimal strategies, whereby the left panel treats the CV case and the right panel the SV case. For both models, the same qualitative conclusions as drawn for the case $T=10$ apply. However, looking at the fraction to be invested in the risky asset, we see that it is considerably higher at the beginning of the investment horizon for $T=30$ compared to $T=10.$ We therefore conclude that it is optimal to invest a higher fraction of the total wealth in equity the younger the saver is as long as particularly the initial wealth-to-contribution ratio is held constant. Conversely, for someone being closer to the retirement date, a more conservative strategy is suggested. The higher expected wealth level simply stems from the longer investment period. After contributing and investing for 20 years, the expected wealth-to-contribution ratio amounts to $\mathbb{E}[P^{\pi^{\star}}_{20}/C_{20}] = 43.02$ and is therefore so high that a significantly lower amount of wealth is invested in equity compared to the case that the savings period only lasts ten years with $P_{0}/C_{0} = 5$. Put differently, if the agent saving 10 years started with a ratio of 43.02, the trading strategies would coincide. The presence of a strictly positive contribution over 30 years can be seen as a high human capital. Starting with a low initial wealth of $P_0=5$ and investing over $30$ years a higher investment ratio can be chosen compared to someone only investing for 10 years.  

In Figure \ref{fig:prolong} we have worked with the parameter values introduced in Section 3.2. Thereby we implicitly assumed that these values remain constant over a time period of 30 years. However, since the parameters might change due to various reasons, it is necessary to investigate whether sticking to a previously computed strategy while parameter values actually changed does highly effect the turnout of the investment. To this end, we consider exemplarily the scenario that the long-term mean of the volatility $\sqrt{\theta}$ rises to $18 \%$ after 15 years. We calculate the optimal strategy when the long-term mean does not change and apply this strategy in a market where there is actually a change. The results are summarized in Table \ref{table_theta}, where the Mean, Variance and CE are listed. The column labeled SVM thereby corresponds to the case that the long-term mean remains at 13\% and the right column shows the result for a market when the long-run mean increases. Obviously, the expected wealth levels nearly coincide, while the variance significantly increases when the strategy is not adjusted for an increased long-term volatility. Likewise, the CE decreases when the strategy is erroneously not adjusted. Hence, when considering such a long-term investment, a fund manager needs to account for changed market parameters; the results in Table \ref{table_theta} show that ignoring such kind of changes may lead to severe disadvantages for the retirement planner. \par
A remark on the absolute value of the variance for the column labeled SVM in Table \ref{table_theta} is due. Comparing this value with the corresponding variance of the optimal terminal wealth for the case $\rho_{\nu} = -0.4$ in Table \ref{table_T}, a sharp increase is observed. A reason for this is the extension of the planning horizon by 20 years and the higher fractional amount of total wealth to be invested in equity. However, this does not mean that an investment over 30 years is disproportionately more risky. This can for instance be seen from a comparison of the respective \emph{coefficient of variation} (CV). The coefficient of variation is defined as the ratio
 of the standard deviation to the mean, that is $\sqrt{\text{Var}[P_{T}^{\pi^{\star}}]} / \mathbb{E}[P_{T}^{\pi^{\star}}].$ As such, it indicates the degree of variability (measured by the standard deviation) in relation to the mean. Calculating the CV for the column $\rho_{\nu} = -0.4$ in Table \ref{table_T} yields about 0.32, and the corresponding value in the SVM case in Table \ref{table_theta} amounts to 0.89.   Hence, the value for the 30-year case is slightly less than three times the value for the 10-year case, so we observe an increase that is approximately proportional to the extension of the investment period. To sum up, a savings period of 30 years leads to a higher amount of total wealth invested in the risky asset, but the overall risk exposure is not disproportionately higher.

\begin{table}
\centering
\begin{tabular}{l| c c }
&\text{SVM} & \text{SVM incr.} $\theta$ \\ \hline \hline
$\mathbb{E}[P^{\pi^{\star}}_{T}]$ 	& 221.09 	&221.18    \\ \hline
$\text{Var}[P^{\pi^{\star}}_{T}]$ 	&38629.79 	&62316.22   \\ \hline 	 
$\text{CE}$ 						& 114.06 	&91.64 	
\end{tabular}
\caption{Expectation, Variance and Certainty Equivalent (CE) when long-term volatility changes to 18\% after 15 years}
\label{table_theta}
\end{table}

\section{Conclusion}

In this paper we study the optimal trading strategy of a Target Date Fund when stochastic volatility is included and random contribution risk is present. 
Using tools from stochastic and numerical optimization we show that the glide path structure that TDFs are frequently identified with is still optimal in such a complex market environment. Given the fact that more and more people use TDFs to save for their pension, our result is of high practical relevance. We show that qualitatively the strategies obtained when presuming a constant volatility do not differ substantially from the ones achieved when stochastic volatility is accounted for, however, the strategies are not simply interchangeable. In the SVM a lower variance of the optimal terminal wealth is generated than in the CVM.  \par
We also outline that a main factor determining the fraction of wealth to be invested in equity is the ratio of initial wealth to initial contribution. Marginally, a low initial wealth induces a more risky strategy. Conversely, a high initial contribution, which is actually the initial value of the accumulated future contributions and therefore allows for the interpretation of human capital, also leads to a riskier investment strategy. As long as the ratio of initial wealth-to-contribution does not change, the strategy is unaffected. \par
It is illustrated that the risk aversion plays an important role for the determination of the optimal trading strategy. A higher risk aversion clearly goes along with a more conservative investment behaviour. Furthermore, the risk aversion determines the shape of the investment strategy plotted as a function of time, i.e., it effects the steepness of the curve. 
Our sensitivity analysis also shows that differences in the drift $\mu_C$ and the volatility $\sigma_C$ of the contribution process only have negligible impacts on the resulting optimal glide path. From a practical point of view, it is therefore possible to offer the same TDF product to a group with homogeneous risk preferences without having perfect knowledge of the contribution dynamics of each individual.

\bibliographystyle{apa}
\bibliography{bibliography}

\end{document}